# Resolving the energy alignment between methylammonium lead iodide and $C_{60}$: an in-situ photoelectron spectroscopy study

Alberto García-Fernández[a,b#*], Karen Radetzky[a,c#], Stefania Riva[a], Birgit Kammlander[a], Brian Rydgren[a,c], Evelyn Johannesson[a,c], Rahul Mahavir Varma[a], Håkan Rensmo[a,c], Ute B. Cappel[a,c*]

[a] Division of X-ray Photon Science, Department of Physics and Astronomy, Uppsala University, Box 516, SE-751 20, Uppsala, Sweden. E-mail: ute.cappel@physics.uu.se

[b] Universidade da Coruña, CICA (Interdisciplinary Center for Chemistry and Biology), Department of Chemistry, Faculty of Science, As Carballeiras, s/n, Campus de Elviña 15071 A Coruña, Spain. E-mail: alberto.garcia.fernandez@udc.es

[c] Wallenberg Initiative Materials Science for Sustainability, Department of Physics and Astronomy, Uppsala University, 751 20 Uppsala, Sweden

# The authors contributed equally to the work.

Understanding and controlling the energy level alignment at interfaces between lead halide perovskites and electron transport layers is crucial for optimizing perovskite-based devices such as solar cells. In this work, we investigated the energy level alignment of $C_{60}$ on in-situ cleaved $MAPbI_3$ single crystals in multiple repeat experiments using photoelectron spectroscopy aiming to resolve inconsistencies reported in earlier studies. Our results show that both materials remain chemically stable upon interface formation, in contrast to the strong reactions typically seen when metals are evaporated on perovskite surfaces. By analyzing the Pb 4f and C 1s binding energies in detail, we determined that $C_{60}$ consistently exhibits a downward energy shift toward $MAPbI_3$, which works against efficient charge extraction. The magnitude of this shift, however, varies between different sample positions, highlighting that small variations in sample surfaces can lead to significant differences in energetic alignment. These findings underscore the decisive role of surface chemistry on interfacial energetics, explain performance variability in perovskite devices, and demonstrate the need to control and accurately measure surface properties. Furthermore, the observed energetic alignment can explain why further interface modification by charge blocking layers or surface passivation is needed for optimized device efficiencies.

## Introduction

In the development of sustainable energy harvesting technologies, perovskite solar cells (PSCs) are a popular candidate. Highly efficient PSCs are based on organic-inorganic metal halide absorber materials.[1] Besides desirable optoelectronic properties such as long carrier diffusion lengths and lifetimes, compositional engineering allows for tuning the material's band gap.[2] Moreover, PSCs are an example of a thin film technology wherein it is possible to fabricate many layers via solution processing.[3] Research in efficiency and stability is progressing along several avenues distinguished by the solar cell architecture, which is divided into n-i-p or p-i-n single junction and multi junction devices. A commonality across these technologies is the need for efficient charge extraction from the perovskite with subsequent charge transport to the contacts.[4] This is achieved by placing the absorber between an electron transport layer (ETL) and a hole transport layer (HTL). The materials are chosen to facilitate selective charge extraction due to the energetic gradient and alignment at the interfaces – with electrons being transferred from the perovskite to the ETL. Many well established ETLs are based on a fullerene motif.[5] In p-i-n devices, functionalized $C_{60}$ and its derivatives are typically employed to tailor the solubility, electronic properties, and interface interactions. For example, PCBM is prevalently used as the ETL in p-i-n devices.[6,7] These devices achieve efficiencies of up to 26%.[8] Moreover, Elnaggar et al. have presented two further fullerene ETLs which improved stability under aging conditions in $MAPbI_3$ based p-i-n devices.[9] Additionally, unaltered $C_{60}$ is also used in photovoltaic devices and has been demonstrated to be beneficial to long-term stability in p-i-n devices based on $Cs_{0.175}FA_{0.750}MA_{0.075}Pb(I_{0.880}Br_{0.120})_3$.[10,11] Momblona et al. demonstrated p-i-n and n-i-p devices employing $C_{60}$ as an electron-selective contact for $MAPbI_3$, with suppressed hysteresis.[12] Nevertheless, subsequent studies have shown that perovskite surface passivation and/or additional interlayers, such as LiF or BCP, can further improve device performance.[13–16] Recently, we have further demonstrated through in situ photoelectron spectroscopy measurements that BCP can modify the energetic landscape at the $C_{60}$ interface through a downward energetic realignment of the $C_{60}$ electronic states, providing a possible explanation for its beneficial effect on charge extraction.[17]

Despite upcoming new approaches, $C_{60}$ layers are commonly fabricated via thermal evaporation.[18] This is advantageous for the assembly of multijunction solar cells, where the deposition parameters of top layers such as solvents and temperature are constricted by the stability of the underlying stack. $C_{60}$ layers are thus also widely applied as an ETL in multijunction perovskite solar cells.[19] Hu et al. utilized $C_{60}$ in double, triple, and quadrupole junction solar cells in contact with mixed cation lead halide perovskites $Cs_{0.1}FA_{0.9}Pb(I_{0.85}Br_{0.15})_3$ and $Cs_{0.05}FA_{0.90}MA_{0.05}PbI_{2.95}Br_{0.05}$ and achieved devices with 28.2%, 28.0%, and 26.9% power conversion efficiency, respectively.[20]

Due to this widespread application, understanding the energetic alignment of $C_{60}$ to the perovskite is of great importance to fully comprehend device performance and limitations.[21–23] Ideally, the LUMO of $C_{60}$ should be below the perovskite conduction band to enable electron transfer. Furthermore, upward energy level shifts (often referred to as band bending) in the $C_{60}$ towards the perovskite would facilitate electron movement away from interface and be beneficial for charge extraction. $MAPbI_3$ is simultaneously a popular perovskite and the first for which a heterojunction with $C_{60}$ was reported.[24] Therefore, numerous studies have already investigated the energetic alignment between $MAPbI_3$ and $C_{60}$, however, without reaching a consensus on the energetics of the interface.

Schulz et al. explored $C_{60}$ layers of increasing thickness on a $MAPbI_3$ thin film substrate with photoelectron spectroscopy and found close alignment of the $C_{60}$ LUMO and the perovskite conduction band.[25] However, Lo et al. reported the band gap of $MAPbI_3$ to be within the $C_{60}$ band gap forming a n-n junction and place the $C_{60}$ LUMO significantly above the perovskite conduction band.[26] A publication by Wang et al. presents a band model with significant upwards band bending towards the interface in the underlying perovskite and downwards band bending in $C_{60}$.[27] This band alignment would be unsuitable for electron extraction from the perovskite to $C_{60}$, even though the $C_{60}$ LUMO is found to be energetically below the $MAPbI_3$ conduction band in their study. For two step perovskite thin film preparation, Shin et al. determined that the energetic alignment to $C_{60}$ depends on the concentration of MAI used in the second step and relate this to the extent of conversion to $MAPbI_3$ and p-doping with excess MAI.[28] Lastly, Quarti et al. presented a theoretical study emphasizing that the surface termination of the $MAPbI_3$ perovskite critically determines the alignment to the $C_{60}$ orbitals and show that beneficial alignment is only present for MAI terminated surfaces.[29] Furthermore, these previous reports show significant differences between the work functions determined for $MAPbI_3$ and for the $C_{60}$, which may depend on the substrate of the perovskite and on differences in sample and surface compositions. These varying reports serve to highlight that the energetic alignment between $MAPbI_3$ and $C_{60}$ is still not fully understood.

Within the study here presented, the energetic alignment between thermally evaporated $C_{60}$ layers and in-situ cleaved $MAPbI_3$ single crystal surfaces is explored for the first time using synchrotron-based soft X-ray photoelectron spectroscopy (PES). As the cleaved single crystal surface is almost free from adsorbates, contaminants, and grain boundaries, we thus base our investigation on a well-defined, clean sample system, in contrast to previous studies. By this carefully designed in-situ investigation, we remove the complexity and variation between different thin film systems. This allows us to successfully isolate the intrinsic chemical and energetic phenomena from external contaminants such as adventitious carbon or oxygen. Furthermore, due to the bulk thickness of the single crystal, the results are not influenced by a substrate below the perovskite. Additionally, we perform a series of duplicate experiments, to be able to study variations in the energy alignment at the perovskite/$C_{60}$ interface. To be able to compare the different experiments, we use relative core level positions: The analysis is based on the energy difference between the Pb 4f and C 1s core level peaks, which are characteristic to $MAPbI_3$ and $C_{60}$, respectively, and can be very accurately determined. This allows for the determination of the energy offset between the two materials and an easy and reliable comparison between different experiments. While we find variations in the energy alignment between different experiments, we find a general trend of a downwards energy shift in the $C_{60}$ layer towards the perovskite and a constant energy offset between the perovskite bands and $C_{60}$ orbitals for thicker $C_{60}$ layers. Understanding these fundamental properties at ideal interfaces is essential for rationalizing and improving functional solar cell architectures, as it enables the scientific community to distinguish whether a particular property is intrinsic to the perovskite/$C_{60}$ system or is generated from external factors.

## Methods

### Single crystal synthesis

Methylammonium lead iodide ($MAPbI_3$) single crystals of approximately 0.5 cm diameter were obtained by inverse temperature crystallization and following the methodology described in our previous publication.[30] In summary, a 1M solution in γ-butyrolactone (GBL) of MAI ($CH_3NH_3I$) (Sigma-Aldrich) and $PbI_2$ (TCI) (1:1) was prepared by stirring at room temperature. Once completely dissolved, the solution was filtered through a 0.45 μm PTFE filter and heated up to 100 °C. After 3 h, selected crystals were collected.

### Photoelectron spectroscopy measurements

PES was carried out during three different beamtimes at the FlexPES beamline at MAX IV.[31] The X-rays from the undulator source were monochromatized using a plane grating monochromator (modified Zeiss SX700). The exit slit was adjusted to reduce the X-ray flux on the sample. Typically, an exit slit of 10 μm was used for a photon energy of 535 eV and either 5 or 10 μm was used for 130 and 758 eV. The endstation employs a Scienta DA30-L(W) spectrometer with a 40 mm MCP/CCD detector. The spectrometer slit was set to 500 μm and a pass energy of 100 eV was used. With these settings and a beamline slit of 10 μm, the combined energy resolution of the spectrometer and the X-rays is 139 meV at 130 eV, 146 meV at 535 eV and 157 meV at 758 eV. The measurement chamber was under ultra-high vacuum conditions with a pressure of $1\times10^{-9}$ to $1\times10^{-10}$ mbar.

Single crystals were mounted on sample plates using two-component conductive epoxy EPO-TEK H20E. The epoxy was cured at 100 °C for 1 h. Clean single crystal surfaces were achieved by blade cleaving under high-vacuum conditions.[30] After initial characterization of the perovskite surface, $C_{60}$ layers were deposited via thermal evaporation with a custom-built evaporator from a molecular source (Sigma Aldrich). The parameters of the evaporations are summarized in Table S1. The samples were characterized before and after each evaporation by core level measurements with different photon energies (535 and 758 eV) and valence band measurements with 130 and 535 eV. For each single crystal, one main spot and at least one control spot were measured.

Within this study, data from three different $MAPbI_3$ single crystals measured on three different beamtimes will be shown and named as: Exp 1 (Data from January 2023), Exp 2 (Data from December 2023), and Exp 3 (Data from September 2024). In addition to measuring all core levels at different photon energies in one spot on the sample (Spot 1), we also measured C 1s and Pb 4f core levels with a photon energy of 535 eV in a different spot (named Spot 2) on the same single crystal and beamtime. As only two core levels were measured, this leads to a reduced X-ray exposure in this spot. Furthermore, during the third experiment, measurements with different X-ray flux densities were carried out by varying the size of the exit slit (Exp 3, Spot 3). Au 4f core level spectra were measured on a gold foil mounted on the sample plate. All core level spectra were energy calibrated by setting the binding energy of Au $4f_{7/2}$ to 84.00 eV. When only this calibration was used, the energy axis in figures is labelled as "Binding Energy". Where we used an internal calibration (for example to the Pb $4f_{7/2}$ core level), the energy axis is labelled as "Relative Binding Energy".

**Data analysis**

Photoelectron core level spectra were fitted with a Pseudo-Voigt function and a Shirley or linear background. The $C_{60}$ coverage and thickness in number of monolayers (ML, 0.7 nm) was calculated based on the peak attenuation of the Pb 4f photoelectron peak and the electron inelastic mean free path ($\lambda$).[32]

The inelastic mean free path of electrons in $C_{60}$ was calculated based on the TPP-2M model.[33] All used parameters can be found in Table S2. The $C_{60}$ thickness in number of monolayers was calculated based on a model which assumes layer-by-layer deposition of $C_{60}$ (Model 1, Fig S1) and is compared to a simpler one, where a layer of uniform thickness is assumed (Model 2, Fig S2). The models though simplified enable us to estimate the film thickness directly from the recorded core level spectra and enable us to relate energetic changes to the distance from the $MAPbI_3/C_{60}$ interface. Both models give comparable results (Fig S3) when the coverage exceeds one monolayer, but Model 1 is expected to be more accurate for coverages below one monolayer and is therefore used when describing the thickness of the evaporated layers. Furthermore, the accuracy of the models is investigated by comparing values calculated with two different photon energies (Fig S4) giving an error in the estimation of about 30%.

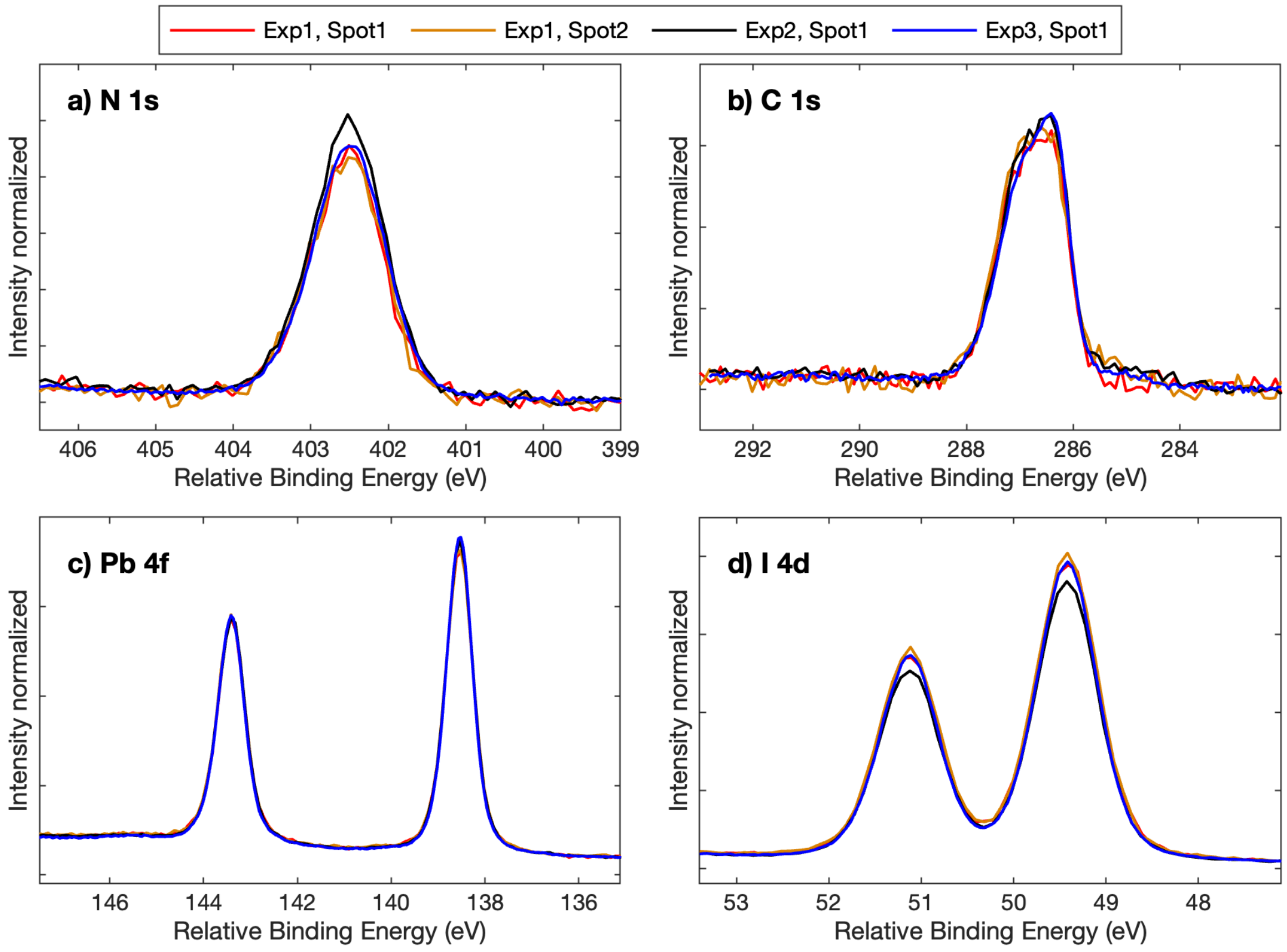

**Fig 1.** Normalized core level spectra of a) N 1s, b) C 1s, c) Pb 4f, d) I 4d of $MAPbI_3$ single crystal surface after cleaving recorded with a photon energy of 535 eV. Data for three different single crystals measured on three different occasions (Exp 1, Exp 2 and Exp 3) and two different measurements spots (Exp 1) on one of the single crystals (Spot 1, Spot 2) is shown. The binding energy scale is calibrated by setting the Pb $4f_{7/2}$ binding energy to 138.54 eV and the intensities are normalized to the Pb 4f intensity. [30]

## Results and discussion

### Core level spectra of $MAPbI_3$ surfaces

Fig 1 shows the core level spectra obtained with a photon energy of 535 eV for surfaces of three different MAPbI3 single crystals. To facilitate comparison, all spectra were energy calibrated to the same Pb $4f_{7/2}$ binding energy of 138.54 eV and normalized to the intensity of the respective Pb 4f core level.[30] Data recorded with a photon energy of 758 eV is shown in Fig S5. The core levels shown represent the different expected components of the $MAPbI_3$ structure, N 1s and C 1s for methylammonium, Pb 4f for $Pb^{2+}$, and I 4d for $I^-$. In all experiments, the shape and intensity of the perovskite core levels are very similar with only very small variations in the relative intensities (see Table S3). In accordance with our previous study, we fitted the C 1s spectrum with two contributions for methylammonium, a surface component at higher binding energies and a bulk component at lower binding energies (Fig S6).[30] The ratios between these two components matches with the ratio found in our previous work (see Table S4) and indicates that the surfaces are mostly, but not fully, terminated by MA (around 70% MA termination). An additional component was included at approximately 285 eV binding energy to model small amounts of adventitious carbon on the surface, which are found in all spectra except for the one obtained for spot 1 in Exp 1. We have estimated the amount of surface contamination quantitatively by calculating the relative contributions of the MA cation and the adventitious carbon to the total C 1s spectrum (Table S4). Herein the adventitious carbon component accounts for less than 5% of the total intensity for all single crystal surfaces. Hence, the surface contamination is very minor.

### Core level spectra after $C_{60}$ evaporation

Following $MAPbI_3$ surface characterization, $C_{60}$ was evaporated in consecutive steps, and the new surface/interface was characterized by PES between each evaporation. Fig 2 shows the measured core level spectra recorded with a photon energy of 535 eV for Exp 1 (spot 1) calibrated to the Fermi level. The same spectra are plotted again in Fig S7, where they are internally energy calibrated to the Pb $4f_{7/2}$ binding energy and normalized to the Pb 4f core level intensity. The same trends as described below were observed at an additional photon energy (758 eV), on another sample spot and for the two other evaporation experiments (Fig S8-S16). Additionally, valence band spectra recorded with a photon energy of 130 eV in Exp 1 are shown in Fig S17. From Fig 2, it can be seen that the intensities of the core levels associated only with the perovskite (N 1s, Pb 4f and I 4d) decrease in intensity. In the C 1s spectra, a new peak appears at approximately 285 eV which increases with consecutive evaporations and is therefore associated with $C_{60}$ deposited on the perovskite surface. As discussed in the SI, the thickness in monolayers of the deposited $C_{60}$ was calculated from the attenuation of the Pb 4f signal compared to the pure perovskite surface. The calculated values show some variation between different photon energies (see Fig S4) and sample spots. The latter could be due to a difference in distance/position relative to the evaporation source.

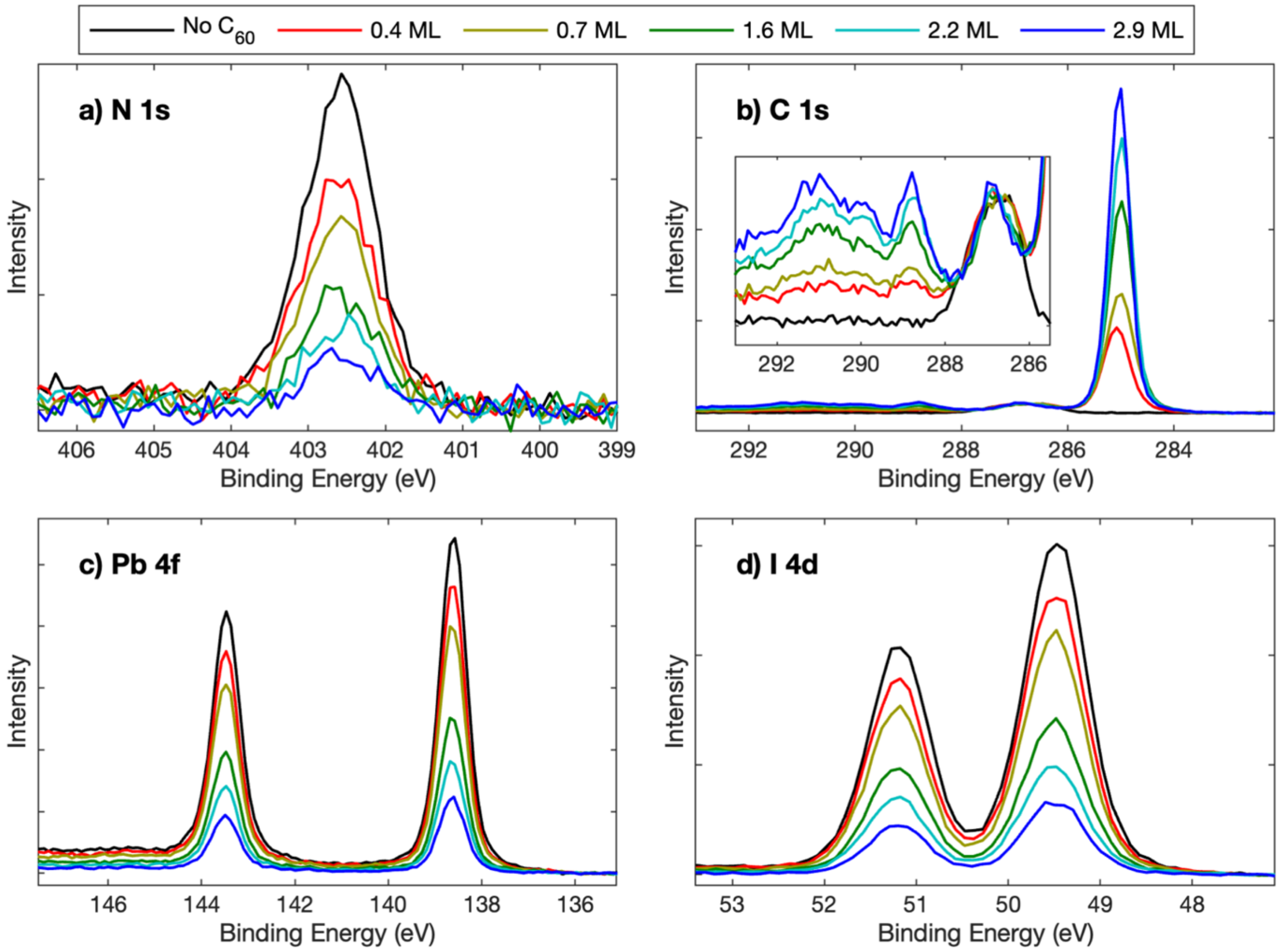

**Fig 2.** Core level spectra of a) N1s, b) C 1s, c) Pb 4f, d) I 4d of a $MAPbI_3$ single crystal surface in Exp 1, spot 1 after cleaving and after consecutive evaporations of $C_{60}$ recorded with a photon energy of 535 eV. The binding energy scale is calibrated to the Fermi level through measurements of the Au 4f core level and the intensities are plotted as measured.

In the Figures below and in the SI, the legend states the values calculated for the respective sample spot with the same photon energy as the shown data.

Insight on the interaction of $C_{60}$ with the perovskite can be gained by comparing the spectra when normalized to the intensity of the perovskite components (Fig S7). It can be observed that the relative intensity and shape of the perovskite components remain very similar upon all repeated evaporations. This confirms that the perovskite surface remains intact upon evaporation of $C_{60}$. This is in contrast to the evaporation of metals onto a perovskite surface, which can cause damage to the perovskite.[34,35] Furthermore, these data indicate that the X-ray irradiation in the experiment does not lead to any significant radiolysis of methylammonium, which has been identified as the main mechanism for X-ray damage of $MAPbI_3$.[36]

The inset in Fig 2b shows the smaller features at binding energies higher than the main $C_{60}$ peak in the C 1s core level. Initially, the $MA^+$ carbon is observed in a binding energy range between 286 and 288 eV. However, additional features are observed, which increase in intensity relative to the perovskite with increased number of $C_{60}$ evaporations. These features are most clear after the final evaporation (2.9 ML) and can be assigned to $C_{60}$ shake-up satellite peaks, which are expected at +1.8, +3.7, +4.8 and +5.9 eV higher binding energies than the main $C_{60}$ peak.[37–39] In Fig 3, we use the C 1s spectra to evaluate the electronic structure of $C_{60}$ after different evaporations. To do so, the spectral intensity is normalized to the intensity of the main $C_{60}$ peak. Furthermore, the spectra are plotted on a relative binding energy axis with the main $C_{60}$ C 1s peak at 0 eV. It can be observed that this main peak is slightly broadened at lower coverage. However, a magnification of the satellite region clearly shows that the satellite features are observed at the expected binding energies for all evaporations with the same relative intensities. This confirms that the $C_{60}$ molecule is intact upon evaporation and that its electronic structure is not significantly changed. Further evidence for this is found in valence band spectra, where the contribution of the perovskite has been subtracted (Fig S17), which clearly show the same peaks after the different evaporations with a small amount of broadening at low coverage.

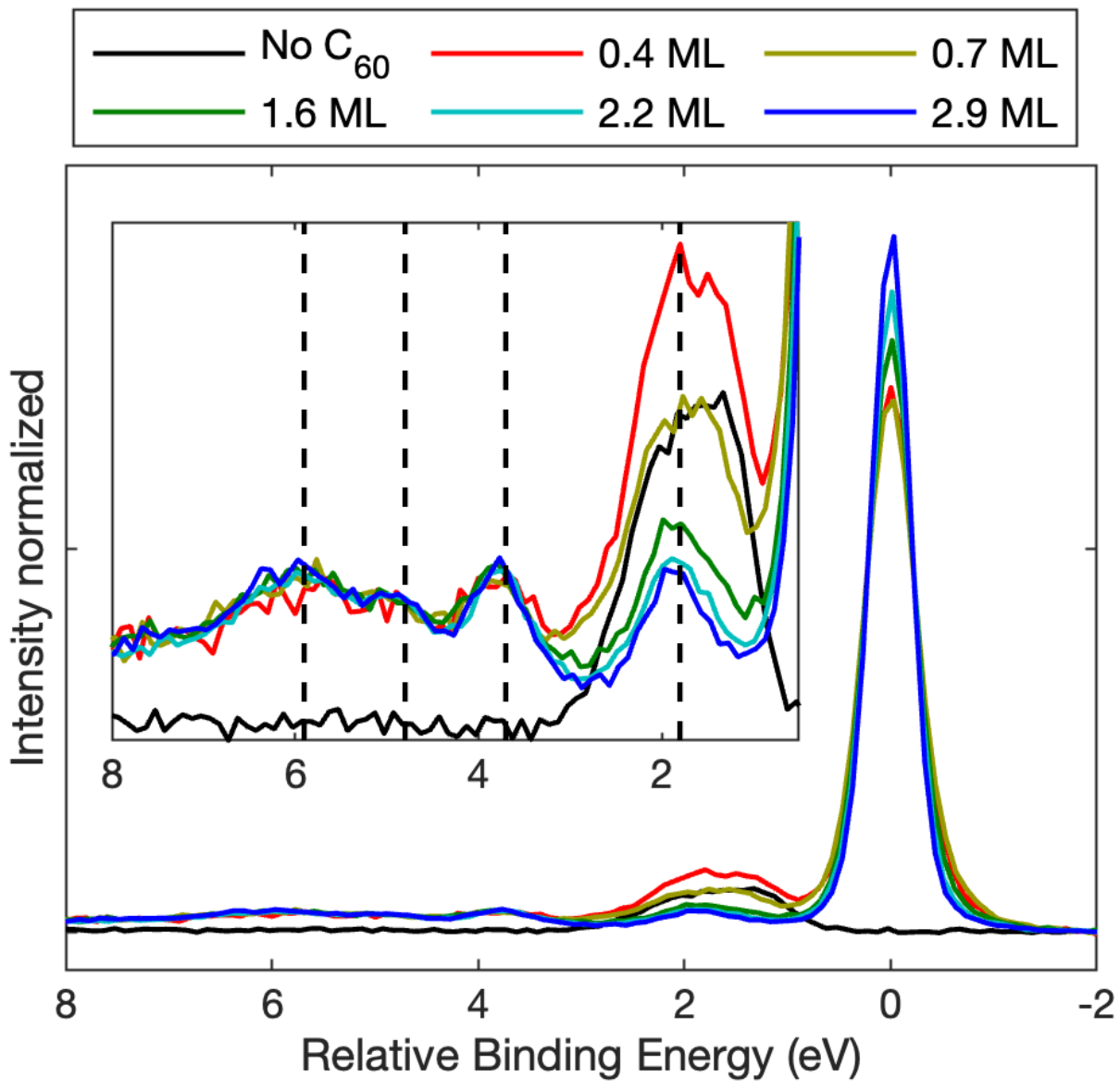

**Fig 3.** C 1s spectra of a $MAPbI_3$ single crystal surface after cleaving in Exp 1, spot 1and after consecutive evaporations of $C_{60}$ recorded with a photon energy of 535 eV relative to the binding energy of the main $C_{60}$ C 1s peak and normalized to its intensity. In the inset, a zoom on the region containing the $C_{60}$ satellite peaks is shown with the expected relative binding energies indicated by dashed vertical lines.

The shape of the C 1s $C_{60}$ spectra can be compared to spectra obtained after evaporation of $C_{60}$ onto a sputtered gold foil (Fig S18). On gold, the $C_{60}$ spectra are considerably broadened and shifted to much lower binding energies at sub-monolayer coverages.[38,39] Furthermore, the satellite peaks have a different shape at low coverage. From the second layer onwards, the spectral shape resembles that of bulk $C_{60}$. The broadening of the first monolayer is due to the chemisorption of the first monolayer on the metal substrate, where the metal's electron density screens the C 1s core hole leading to a different final state energy and therefore a lower binding energy. Such effects have been observed for $C_{60}$ on different metals but should not be observed when $C_{60}$ is adsorbed on a semiconductor.[38–42] The initial broadening of the $C_{60}$ main peak on $MAPbI_3$ could be interpreted in two ways: interaction between the perovskite and $C_{60}$ or a broadening at coverages less than one monolayer. Overall, this indicates that there is a different interaction between $C_{60}$ and $MAPbI_3$ than in between $C_{60}$ molecules, however the effect of this interaction on the electronic structure of $C_{60}$ is small.

Given that both perovskite and $C_{60}$ remain intact during evaporation and that there are only minor changes in the electronic structure, the shifts in core level binding energies can be used to determine the energy alignment at the $MAPbI_3$/ $C_{60}$ interface based on a rigid band model.[43] In this model, the energy difference between core levels and valence and conduction band is assumed to be constant for a given material and core level binding energies can thus be used to follow changes in the band alignment. For the perovskite, this effect can also be observed in overlapping core levels for the N 1s and I 4d core levels after evaporation when internally calibrated to the Pb $4f_{7/2}$ binding energy (Fig S7). In the following, we will focus on using the Pb $4f_{7/2}$ binding energy of the perovskite and the C 1s binding energy of the main $C_{60}$ peak to follow the band alignment at the perovskite / $C_{60}$ interface.

Fig 4a and b show the binding energies of the Pb $4f_{7/2}$ peak of the perovskite and the C 1s peak of the $C_{60}$ calibrated to the Fermi level (as described in the experimental section) during the consecutive evaporations determined from measurements in two different spots on the sample and at two different photon energies (535 eV, 758 eV) in Exp1. It can be seen that the Pb $4f_{7/2}$ binding energy is quite constant (shifts of less than 0.05 eV for different evaporations within one measurement set). However, some variations for the different spots and photon energies are observed with peak positions shifting to higher binding energies upon addition of more $C_{60}$ in spot 1 and to lower binding energies in spot 2. For the C 1s core level of $C_{60}$, a shift to lower binding energy is observed with increasing amount of $C_{60}$. However, the absolute binding energies vary between different spots, and the observed shift is more gradual in spot 2. Overall, the differences in trends and absolute binding energy position highlight the need for repeated experiments. Differences in absolute binding energies can stem from differences in Fermi level position but also from sample charging effects under X-ray illumination.

To get further insight, we used a data analysis method for comparing the three different $C_{60}$ evaporation experiments on $MAPbI_3$, (Exp 1, 2 and 3), which is based on the energy difference between the $C_{60}$ C 1s peak and the perovskite Pb $4f_{7/2}$ peak (described in the next section). This energy difference in shown in Fig 4c for Exp 1. In this figure, it is very clear that the data obtained at two different photon energies overlap, while the data from a 2nd spot on the sample shows a different value. In one of the experiments (Exp 3), we observed some sample charging with the Pb $4f_{7/2}$ core level being at higher binding energy than in previous experiments (Fig S13-S15). To estimate the effect of sample charging on our analysis, the substrate and overlayer system was measured at different incident X-ray flux values (Fig S16). While the measured absolute binding energy of the Pb 4f and C 1s core levels increases as a function of incident photon flux, the energetic difference between the core levels remains constant (Fig S19). Hence, we conclude that charging events do not impact the presented analysis. In contrast, such effects will impact an analysis based on absolute binding energies, such as valence band edge determinations and secondary electron cut-off measurements to determine the work function.

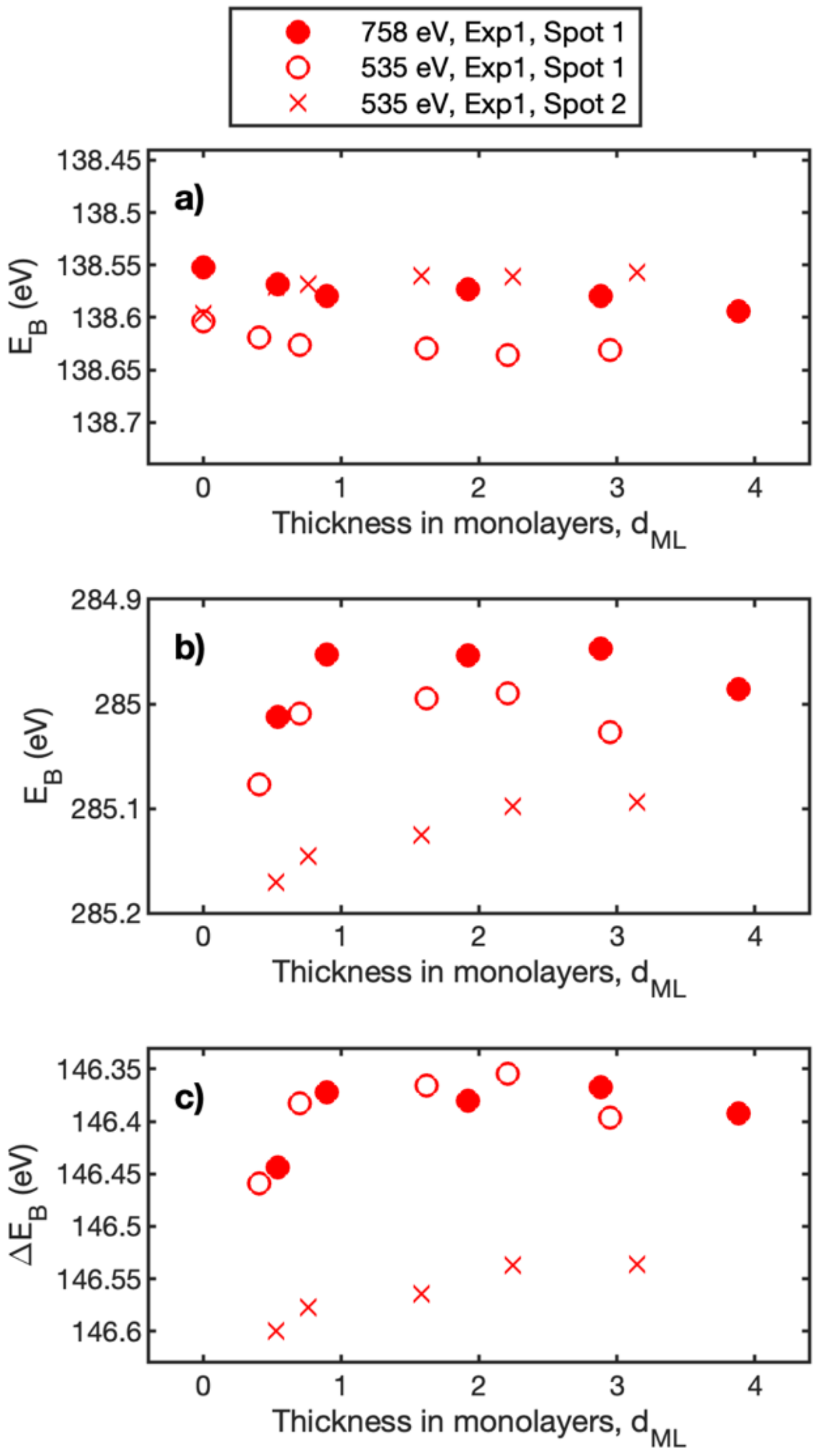

**Fig 4.** Peak binding energy calibrated to the Fermi level of a) Pb $4f_{7/2}$ (perovskite peak) and b) C 1s ($C_{60}$ peak) after different evaporations of $C_{60}$, determined from Exp1 in 2 different spots on the sample and at two different photon energies (535 eV and 758 eV). c) Energy difference ($\Delta E_B$) between the between the binding energies in a) and b).

**Energy alignment**

To be able to compare different $C_{60}$ evaporation experiments, we focused on determining an energy offset parameter between the $C_{60}$ HOMO level and $MAPbI_3$ valence band (VB) for the different evaporations. In this way, the energy offset can be plotted versus the $C_{60}$ thickness in monolayers (determined from the Pb 4f core level intensity as described in the experimental and in the SI). To calculate the energy offset, the binding energies of the Pb $4f_{7/2}$ (perovskite) and the C 1s ($C_{60}$) recorded after evaporation were used. Additionally, the energy differences between the core level and the HOMO level / valence band edge need to be known. We carried out the determination of the perovskite valence band maximum (VBM) and $C_{60}$ HOMO level using material-specific analysis methods to account for the distinct electronic structures of the system (Fig S20-S23). For halide perovskites, the valence band edge was extracted using a logarithmic representation of the photoemission onset (Figure S20). This approach is necessary due to the low density of states near the band edge and the location of the VBM away from the Brillouin zone center, which leads to a gradual onset that is not accurately captured by linear extrapolation. Previous studies have shown that linear methods tend to overestimate the VBM position,[44,45] whereas logarithmic analysis provides values consistent with single-crystal ARPES measurements and a more realistic description of the band edge.[46] By a combination of core level and valence band measurements, which were energy calibrated internally, we then determined the energy difference between the Pb $4f_{7/2}$ binding energy and the valence band edge to be 137.39 eV (Fig S21). This value is very similar to values found previously for mixed cation lead iodide perovskites thin films but somewhat higher than a value determined for $MAPbI_3$ thin films.[47,48]

For $C_{60}$ the HOMO onset was determined using linear extrapolation (Fig S23), following the established convention for organic semiconductors, where the more localized electronic states give rise to a well-defined leading edge.[22,49] This combined approach ensures a consistent and physically meaningful determination of the energy level alignment across the hybrid interface. The energy difference between C 1s and the HOMO level for $C_{60}$ was determined to be 283.24 eV (Fig S23). From these energy differences together with the Pb $4f_{7/2}$ and C 1s binding energies, the energy offset between the perovskite valence band and the $C_{60}$ HOMO level ($\Delta E_{HOMO-VB}$) can be determined according to the following equation:

$$\Delta E_{HOMO-VB} = (E_{C\,1s} - 283.24\ eV) - (E_{Pb\,4f_{7/2}} - 137.39\ eV).$$

Fig 5a shows a schematic representation of how $\Delta E_{HOMO-VB}$ is calculated. Note that a positive $\Delta E_{HOMO-VB}$ signifies that the $C_{60}$ HOMO level is below the perovskite valence band in energy (as indicated in Fig 5a). Fig 5b shows the values of $\Delta E_{HOMO-VB}$ at different monolayer thicknesses $d_{ML}$ for all three experiments in all measurement spots. All calculated values are positive and therefore indicate hole blocking properties of $C_{60}$ at the interface to the perovskite, which is desirable for the function of solar cells. Measurements in the same spot with 535 and 758 eV photon energy give similar energy offset values but appear at slightly different points on the x-axis which can be explained by the uncertainty in the determined thicknesses (Fig S4). However, the data exhibits an energetic spread at low coverages below 3 ML of up to 0.2 eV between different spots and across Exp 1 and 2 showing that the energy offset at low coverages is not always the same. Differences in the accumulated X-ray exposure between measurement spots may also have a small contribution to the observed variability; however, no characteristic signatures of $MAPbI_3$ radiolysis, such as degradation of the $MA^+$ component, were observed under the measurement conditions used here, suggesting that irradiation-induced degradation is not the dominant source of the observed energy-level variations.[36,50]

Quarti et al. suggested that the surface termination can impact the energetic alignment of $C_{60}$ on $MAPbI_3$ as demonstrated through calculations.[29] A similar dependence on the surface termination has also been previously reported for the $CsPbBr_3$/ $C_{60}$ interface.[23] Loh et al. determine a 0.4 eV shift of the molecular orbitals between CsBr and $PbBr_2$ termination. In the present case, the core level spectra of the cleaved $MAPbI_3$ surfaces at both 535 eV and 758 eV show very similar relative core level intensities (Figure 1, Figure S5 and Table S3). We therefore do not see any evidence for significant differences in the surface termination between spots and samples, as these should change the relative core level intensities.[51] In particular the nitrogen and carbon intensities relative to lead should change, as MAI termination would significantly increase the amount of probed MA compared to $PbI_2$ termination given the surface sensitivity of the measurement. Such effects are not observed (Table S3) supporting our conclusion that the surface termination is similar in all measured samples. However, we note that spot 1 in Exp 1 was cleaner than the other measurement spots without the presence of any adventitious carbon (Fig 1, Fig S5-6 and Table S4). This spot shows a lower value of the energy offset at low coverages compared to all other measurement spots. This suggests that in addition to surface termination, also the presence of surface contaminants can influence the energy alignment between $MAPbI_3$ and $C_{60}$. Yet, we find no trend relating the amount of contamination to the strength of the offset, only a distinct difference between the clean spot to all others. Despite the observed spread in the initial alignment, the values for $\Delta E_{HOMO-VB}$ become smaller at higher values of $d_{ML}$ in all measurement spots in all experiments. This general trend indicates a downward energy shift in the $C_{60}$ HOMO level towards the interface with $MAPbI_3$. Furthermore, at higher $C_{60}$ coverages (> 4.5 ML), the energy offset between the $MAPbI_3$ valence band and the $C_{60}$ HOMO reaches approximately 0.52 eV for Exp 3 (maximum deviation of 0.03 eV). In the contamination-free Exp 1 spot 1 measurements, a similar average offset of 0.53 eV is already reached at thicknesses above 1 ML (maximum deviation of 0.02 eV). These values are significantly larger compared to 0.10 eV reported by Lo et al.[26] while good agreement is present with Wang et al. who report 0.57 eV.[27]

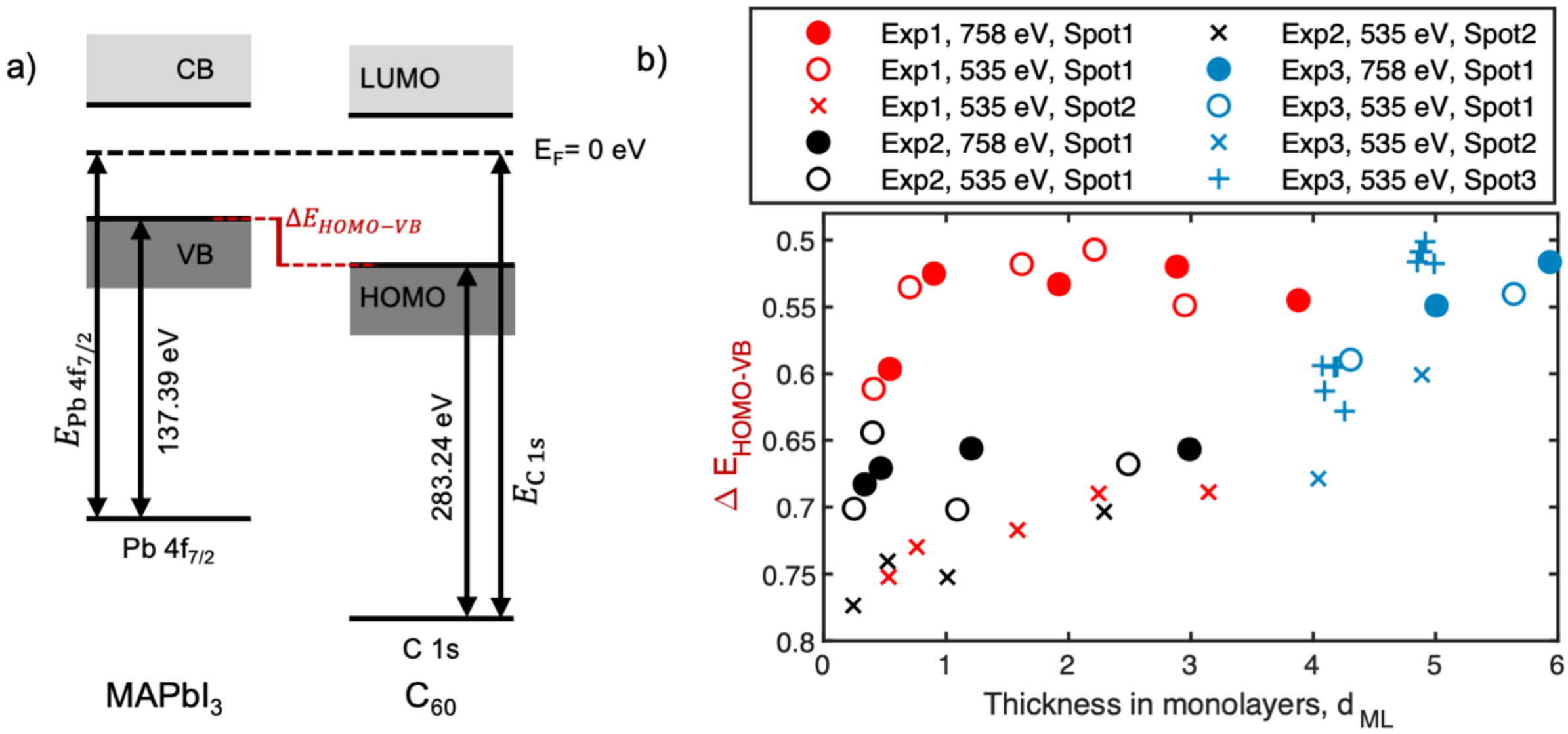

**Fig 5.** a) Schematic representation illustrating the methodology of how $\Delta E_{HOMO-VB}$ is defined and calculated. b) Calculated $\Delta E_{HOMO-VB}$ plotted vs. $C_{60}$ thickness in monolayers, $d_{ML}$, determined from Pb 4f and C 1s core level measurements after $C_{60}$ evaporation on MAPbI3 single crystal surfaces. The different colors indicate three different evaporation series on three different $MAPbI_3$ single crystals. The different symbols indicate different photon energies (535 eV and 758 eV) and measurement spots on the sample.

Based on our observations, we construct an energy diagram for the $MAPbI_3/C_{60}$ interface (Fig 6). In this diagram, we also include the conduction band edge of $MAPbI_3$ at 1.57 eV above the valence band edge based on the band gap ($E_g$) of $MAPbI_3$.[52] Additionally, the LUMO of $C_{60}$ is calculated in the same way.

However, for $C_{60}$, different HOMO-LUMO gaps have been applied to construct energy diagrams. Some published energy diagrams are based on the optical band gap which typically lies between 1.7 eV and 1.9 eV.[53–57] The range of possible LUMO values based on the optical band gap values are indicated in Fig 6 by a grey shaded area. However, the optical band gap includes excitonic contributions giving the energy of a bound exciton rather than of a free electron. Other reports thus used a band gap of 2.3 eV determined via a combination of UPS and IPES,[25,26,58] also referred to as the mobility gap.[59] Schulz et al. demonstrated that this HOMO-LUMO separation does not depend on the $C_{60}$ layer thickness.[25] As this is supported by our own data which shows only minimal electronic interaction between $C_{60}$ and the underlying $MAPbI_3$, we apply a HOMO-LUMO separation of 2.3 eV in our energy diagram at all distances from the interface to indicate the LUMO position based on the mobility gap (Fig 6, blue line).

Compared to the optical band gap, this places the LUMO of $C_{60}$ above the conduction band edge of the perovskite, which would create an energetic barrier for electron transport away from the $C_{60}/MAPbI_3$ interface. At the highest $C_{60}$ thicknesses investigated here, the offset between the LUMO and the conduction band is -0.19 eV. At low thicknesses, the LUMO is below the $MAPbI_3$ conduction band for some of the surfaces.

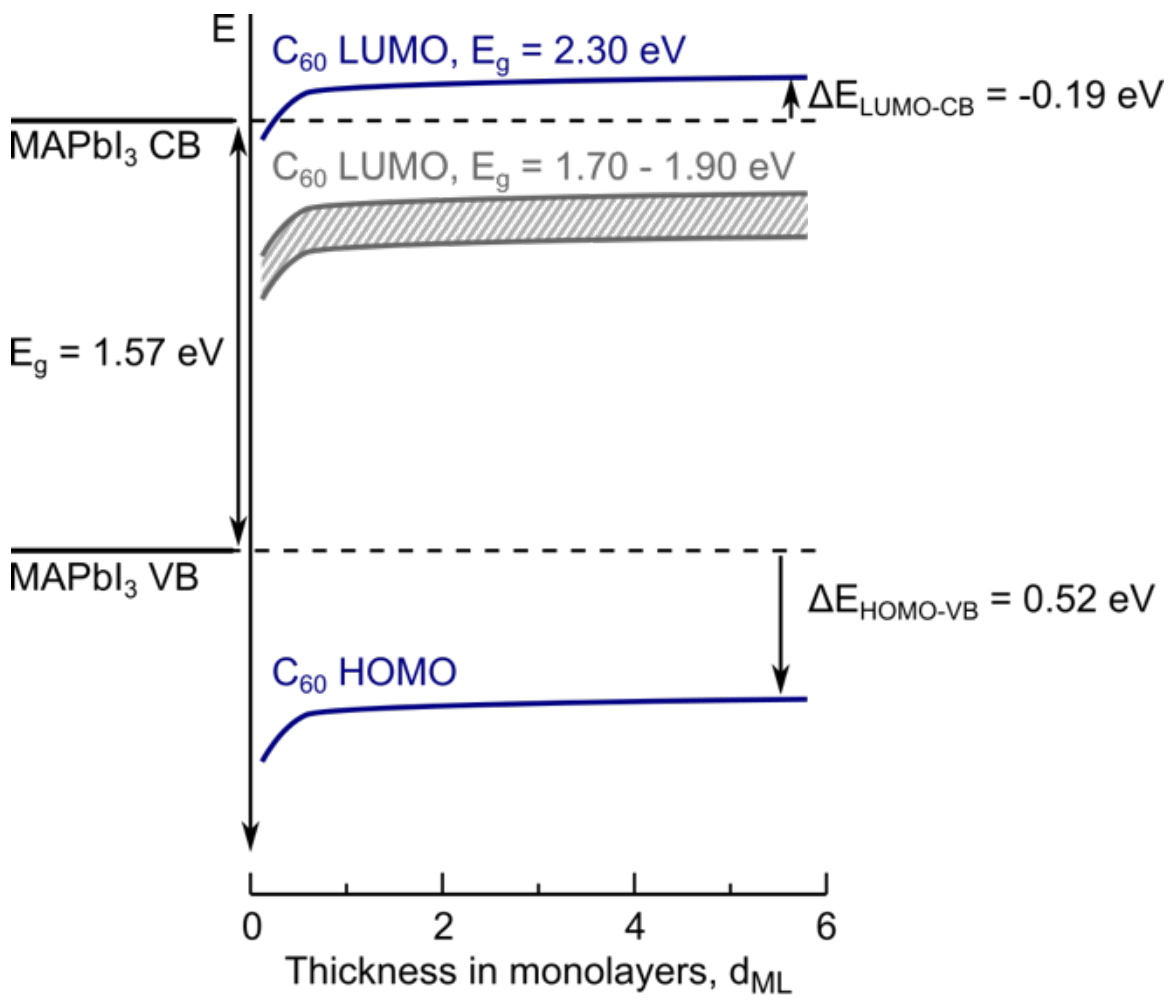

**Fig 6.** Energy diagram of the $MAPbI_3/C_{60}$ interface constructed based on the data in Fig 5b and band gap values for $MAPbI_3$ and for $C_{60}$ as discussed.

The downward energy shift observed upon $C_{60}$ deposition is consistent with interfacial charge redistribution and electron transfer from $MAPbI_3$ to $C_{60}$. Previous photoelectron spectroscopy studies of $C_{60}/MAPbI_3$ interfaces have reported an interface dipole and $C_{60}$ band bending indicative of electron transfer from the perovskite to $C_{60}$.[27] This interpretation is also supported by theoretical studies predicting charge transfer from $MAPbI_3$ to $C_{60}$ at the interface.[29,60] Due to our in-situ approach and sequential deposition of $C_{60}$, we are also able to follow the evolution of the chemical structure of the perovskite and $C_{60}$ through details in the core and valence band spectra (Figure 3, S7 and S17). As discussed above, we do not observe significant changes in the perovskite core-level spectra or chemical components upon $C_{60}$ deposition that would indicate substantial surface degradation or changes in surface composition. We therefore consider changes in the perovskite surface chemistry unlikely to be the primary origin of the observed energetic shift. Likewise, the absence of significant changes in the spectral features associated with the C 1s of $C_{60}$ and perovskite surface suggests that there is no substantial orbital hybridization between $C_{60}$ and perovskite states. Instead, the observed energetic evolution is consistent with charge redistribution upon contact formation, suggesting the presence of an interfacial dipole and band bending within the $C_{60}$ layer. Similar charge-induced energetic realignment has been experimentally observed in our previous in-situ photoelectron spectroscopy studies of TIPS-pentacene/$MAPbI_3$ interfaces.[61] The precise magnitude of the charge transfer may depend on the local surface chemistry and defect environment; however, our results point to interfacial charge redistribution as the most plausible origin of the observed downward band bending.

As $C_{60}$ is used successfully as an electron selective contact in perovskite solar cells, these discrepancies can be interpreted in the following ways:

1. The LUMO energy determined by applying a 2.3 eV band gap lies higher than other states that can accept electrons via charge injection.
2. In optimized devices based on solution-processed or evaporated perovskites, interface-specific effects may lead to a more favorable energetic alignment than that predicted from idealized band positions
3. Under light illumination, ion migration leads to a rearrangement of the energy alignment to the interface and makes it more favorable.[62]
4. the additional layers in a device, such as charge blocking layers and metal contacts lead to a further modification of the energetic landscape, which enables efficient charge extraction. We have for example recently shown that evaporation of BCP onto $C_{60}$ can lead to a downward energy shift of the $C_{60}$, which would make its alignment favorable again for charge extraction in p-i-n perovskite solar cells.[17]

Overall, our results indicate that the energy alignment between $MAPbI_3$ and $C_{60}$ may vary and is very sensitive to the exact surface structure, and in the presented case can likely be influenced by a variation in very small amounts of surface contamination or defect states. Further variations can be introduced by differences in surface termination.[23,29] These results explain the spread of values found in literature to some degree. In particular, when using multicrystalline thin films of $MAPbI_3$, the surface chemistry of the $MAPbI_3$ can vary significantly with the exact preparation procedure and environment. Furthermore, $C_{60}$ shows some downward energy shifts towards the perovskite, which could create an energetic barrier for electron transport away from the $C_{60}/MAPbI_3$ interface.

The single-crystal measurements presented here provide a well-defined model system for isolating the effects of surface chemistry and interfacial interactions on energy-level alignment, while minimizing the additional complexity associated with grain boundaries and heterogeneous film morphology. In practical polycrystalline perovskite films, however, the interfacial energetics are expected to be more heterogeneous. Grain boundaries, surface roughness, local variations in composition and defect density, and differences in film preparation and processing can modify the local electrostatic potential and chemical environment at the perovskite/transport-layer interface. Consequently, the absolute energy shifts measured on an ideal single-crystal surface should not necessarily be considered representative of every location in a polycrystalline film. Rather, the single-crystal results provide a controlled reference for identifying the fundamental interfacial mechanisms that can contribute to the energetic variability observed in practical devices. In this context, the interfacial effects identified here therefore establish a baseline for understanding the energetic heterogeneity of polycrystalline devices, to which morphology and processing introduce further variability.

## Conclusions

In conclusion, we investigated the energy alignment of $C_{60}$ on cleaved $MAPbI_3$ single crystal surfaces in several in-situ experiments. We find that both $MAPbI_3$ and $C_{60}$ remain intact upon interface formation unlike what is observed for evaporation of metals on perovskite surfaces. Furthermore, our results indicate some interaction between thin layers of $C_{60}$ and the perovskite, but no significant changes to the electronic structure.

By determining the thickness of the evaporated $C_{60}$ layer from the Pb 4f core level intensity attenuation and the energy offset between the MAPbI3 valence band and the $C_{60}$ HOMO level from relative core level binding energies, we are able to compare a series of different repeat experiments. We find that $C_{60}$ shows a downward energy shift toward MAPbI3 in repeat experiments, similar to what others have reported previously. However, the magnitude of the energy shift varied between different experiments. When the coverage of $C_{60}$ reached several monolayers, the energy offset between the HOMO and the valence band became 0.52 eV in some of the experiments. A HOMO-LUMO gap of 2.3 eV suggests that the $C_{60}$ LUMO is above the $MAPbI_3$ conduction band by 0.19 eV in these cases, i.e. too high for charge extraction. This value as well as the unfavorable energy shift could explain why further interface modification, for example by charge blocking layers, is needed for efficient solar cells when using $C_{60}$ as an ETL. Overall, our study highlights that energy alignment directly at the interface in between perovskites and electron transport layers can vary between different repeat experiments, even for single crystals, and therefore is very sensitive to the exact surface composition. This means that the energy alignment for a given system has to be determined carefully. The energy diagram presented here therefore serves as reference for which energy alignment can be expected at an intrinsic perovskite / $C_{60}$ interface in absence of grain boundaries.

## Conflicts of interest

There are no conflicts to declare.

## Supporting Information

The supporting information contains details related to the employed thickness calculations, additional spectra, and valence band edge determination.

## Acknowledgements

We acknowledge the MAX IV Laboratory for beamtime on the FlexPES beamline under proposal 20220600, 20230156, and 20240429. Research conducted at MAX IV, a Swedish national user facility, is supported by Vetenskapsrådet (Swedish Research Council, VR) under contract 2018-07152, Vinnova (Swedish Governmental Agency for Innovation Systems) under contract 2018-04969 and Formas under contract 2019-02496.

We thank Alexei Preobrajenski, Stephan Appelfeller, and Alexander Generalov for their support during the beamtime. This work was partially supported by the Wallenberg Initiative Materials Science for Sustainability (WISE) funded by the Knut and Alice Wallenberg Foundation. We thank the Swedish Research Council (Grant numbers 2022-03168 and 2023-05072) for funding. A.G.-F acknowledges support from a Beatriz Galindo junior fellowship (BG23/00033) from the Spanish Ministry of Science and Innovation. S.R. and H.R. acknowledge support from the Swedish Energy Agency (P50626-1).

# Supporting Information for:

# Resolving the energy alignment between methylammonium lead iodide and $C_{60}$: an in-situ photoelectron spectroscopy study

Alberto García-Fernández, Karen Radetzky, Stefania Riva, Birgit Kammlander, Brian Rydgren, Evelyn Johannesson, Rahul Mahavir Varma, Håkan Rensmo, Ute B. Cappel

Table S1. Parameters used for the thermal evaporation of $C_{60}$ in all experiments. A custom built evaporator was attached directly to the preparation chamber at the end station.

| Experiment | Evaporation | Current [A] | Voltage [V] | Time [min] | Chamber pressure [mbar] |
|---|---|---|---|---|---|
| 1 | 1 | 6.1 | 1.0 | 3 | $1.0\times 10^{-8}$ |
| 1 | 2 | 6.1 | 1.0 | 2 | 7-9$\times 10^{-9}$ |
| 1 | 3 | 6.1 | 1.0 | 6 | $1.0\times 10^{-8}$ |
| 1 | 4 | 6.1 | 1.0 | 6 | $7.0\times 10^{-8}$ |
| 1 | 5 | 6.1 | 1.0 | 12 | $8\times 10^{-9}$ |
| 2 | 1 | 6.0 | 0.9 | 3 | $6.0\times 10^{-8}$ |
| 2 | 2 | 6.0 | 0.9 | 5 | $3.7\times 10^{-8}$ |
| 2 | 3 | 6.3 | 1.1 | 8 | $2.0\times 10^{-8}$ |
| 2 | 4 | 6.5 | 1.1 | 10 | $1.3\times 10^{-8}$ |
| 3 | 1 | 6.0 | 1.3 | 6 | $2.0\times 10^{-8}$ |
| 3 | 2 | 6.0 | 1.3 | 3 | $4\times 10^{-9}$ |

Table S2. Parameters used in the inelastic mean free path calculations.[1,2]

| Core level | Photon energy [eV] | $E_{kin}$ [eV] | $C_{60}$ density [g/cm$^3$] | $C_{60}$ molar mass [g/mol] | $C_{60}$ valence electrons | $C_{60}$ band gap | IMFP [nm] |
|---|---|---|---|---|---|---|---|
| Pb 4f | 535 | 396 | 1.65 | 720.66 | 240 | 1.86 | 1.35 |
| Pb 4f | 758 | 619 | 1.65 | 720.66 | 240 | 1.86 | 1.87 |
| Au 4f | 535 | 451 | 1.65 | 720.66 | 240 | 1.86 | 1.48 |

Table S3. Intensity ratios for the core level spectra shown in Figure 1 recorded with 535 eV on the cleaved $MAPbI_3$ single crystal surfaces corrected for photoionization cross sections.

| Experiment, spot | I 4d / Pb 4f | C 1s(total MA)/ Pb 4f | N1s/ Pb 4f |
|---|---|---|---|
| Exp 1, spot 1 | 4.57 | 1.61 | 1.15 |
| Exp 1, spot 2 | 4.66 | 1.73 | 1.15 |
| Exp 2, spot 1 | 4.28 | 1.70 | 1.31 |
| Exp 3, spot 1 | 4.43 | 1.60 | 1.19 |

Table S4. Fitted peak intensities of the C 1s peaks associated with MA and adventitious carbon divided by the total C 1s signal to assess the extent of surface contamination determined from core level fitting the spectra recorded on the cleaved $MAPbI_3$ single crystal surfaces with a photon energy of 535 eV. The ratios between the MA surface and bulk contributions are also given.

| Experiment, spot | C 1s(MA bulk) /C 1s(total) [%] | C 1s(MA surface) /C 1s(total) [%] | C 1s(adv. Carbon) /C 1s(total) [%] | C 1s(MA surface) / C 1s(MA bulk) |
|---|---|---|---|---|
| Exp 1, spot 1 | 57.7 | 42.3 | 0.00 | 0.73 |
| Exp 1, spot 2 | 53.8 | 42.5 | 3.70 | 0.79 |
| Exp 2, spot 1 | 57.2 | 38.1 | 4.75 | 0.67 |
| Exp 2, spot 2 | 57.4 | 38.0 | 4.61 | 0.66 |
| Exp 3, spot 1 | 56.1 | 42.2 | 1.73 | 0.75 |
| Exp 3, spot 2 | 59.4 | 36.4 | 4.22 | 0.61 |

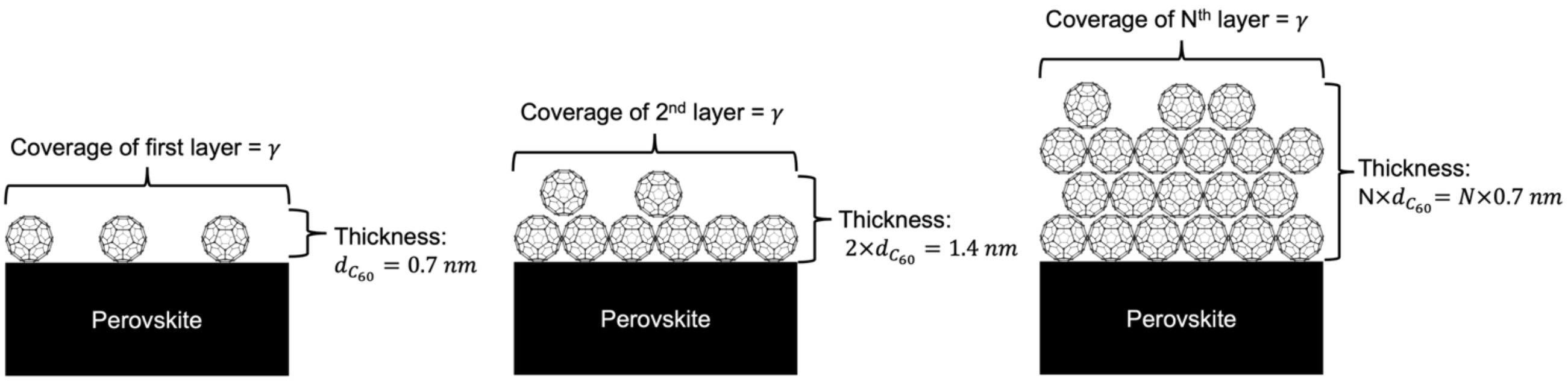

$$I = I_0(1-\gamma) + I_0\gamma\exp(-\frac{d_{C_{60}}}{\lambda})$$

$$\gamma = \frac{\frac{I}{I_0} - 1}{\exp\left(-\frac{d_{C_{60}}}{\lambda}\right) - 1}$$

where $0 < \gamma \leq 1$

Thickness in monolayers: $d_{ML} = \gamma$

$$I = I_0(1-\gamma)\exp(-\frac{d_{C_{60}}}{\lambda}) + I_0\gamma\exp(-\frac{2\times d_{C_{60}}}{\lambda})$$

$$\gamma = \frac{\frac{I}{I_0} - \exp\left(-\frac{d_{C_{60}}}{\lambda}\right)}{\exp\left(-\frac{2\times d_{C_{60}}}{\lambda}\right) - \exp\left(-\frac{d_{C_{60}}}{\lambda}\right)}$$

where $0 < \gamma \leq 1$

Thickness in monolayers: $d_{ML} = \gamma + (2 - 1)$

$$I = I_0(1-\gamma)\exp\left(-\frac{(N-1)d_{C_{60}}}{\lambda}\right) + I_0\gamma\exp(-\frac{N\times d_{C_{60}}}{\lambda})$$

$$\gamma = \frac{\frac{I}{I_0} - \exp\left(-\frac{(N-1)d_{C_{60}}}{\lambda}\right)}{\exp\left(-\frac{N\times d_{C_{60}}}{\lambda}\right) - \exp\left(-\frac{(N-1)d_{C_{60}}}{\lambda}\right)}$$

where N is chosen so that $0 < \gamma \leq 1$

Thickness in monolayers: $d_{ML} = \gamma + (N - 1)$

Figure S1: Model 1 for calculating the thickness in number of monolayers of $C_{60}$. The model is based on the assumption that the $C_{60}$ layer on the surface forms layer-by-layer[3] and that the thickness of one monolayer of $C_{60}$ ($d_{C_{60}}$) is 0.7 nm (diameter of $C_{60}$).[4] $I_0$ describes the intensity of a given perovskite core level (Pb 4f was used for the calculations in this paper) without $C_{60}$ on top, while $I$ is the intensity of the same core level measured with the same experimental settings after deposition of $C_{60}$. $\lambda$ is the inelastic mean free path (IMPF) of the photoelectron.

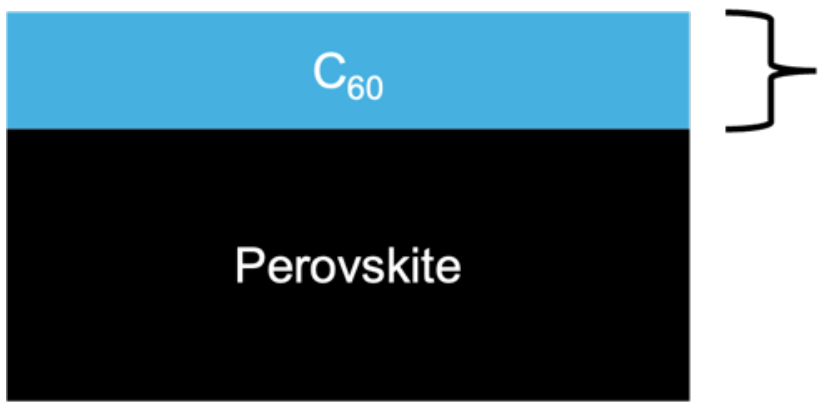

$$I = I_0\exp(-\frac{d}{\lambda})$$

$$d = -\lambda\ \ln(\frac{I}{I_0})$$

Thickness in monolayers $d_{ML} = \frac{d}{d_{C_{60}}} = \frac{d}{0.7\ nm}$

Figure S2: Model 2 for calculating the thickness of $C_{60}$. The model is based on the assumption that the $C_{60}$ layer has a uniform thickness of *d*. The thickness in number of monolayers is then obtained by dividing d by the diameter of $C_{60}$ (0.7 nm). While this model gives similar values to the one above, a uniform layer thickness is not physically possible for layers below one monolayer. $I_0$ describes the intensity of a given perovskite core level (Pb 4f was used for the calculations in this paper) without $C_{60}$ on top, while $I$ is the intensity of the same core level measured with the same experimental settings after deposition of $C_{60}$. $\lambda$ is the inelastic mean free path (IMPF) of the photoelectron.

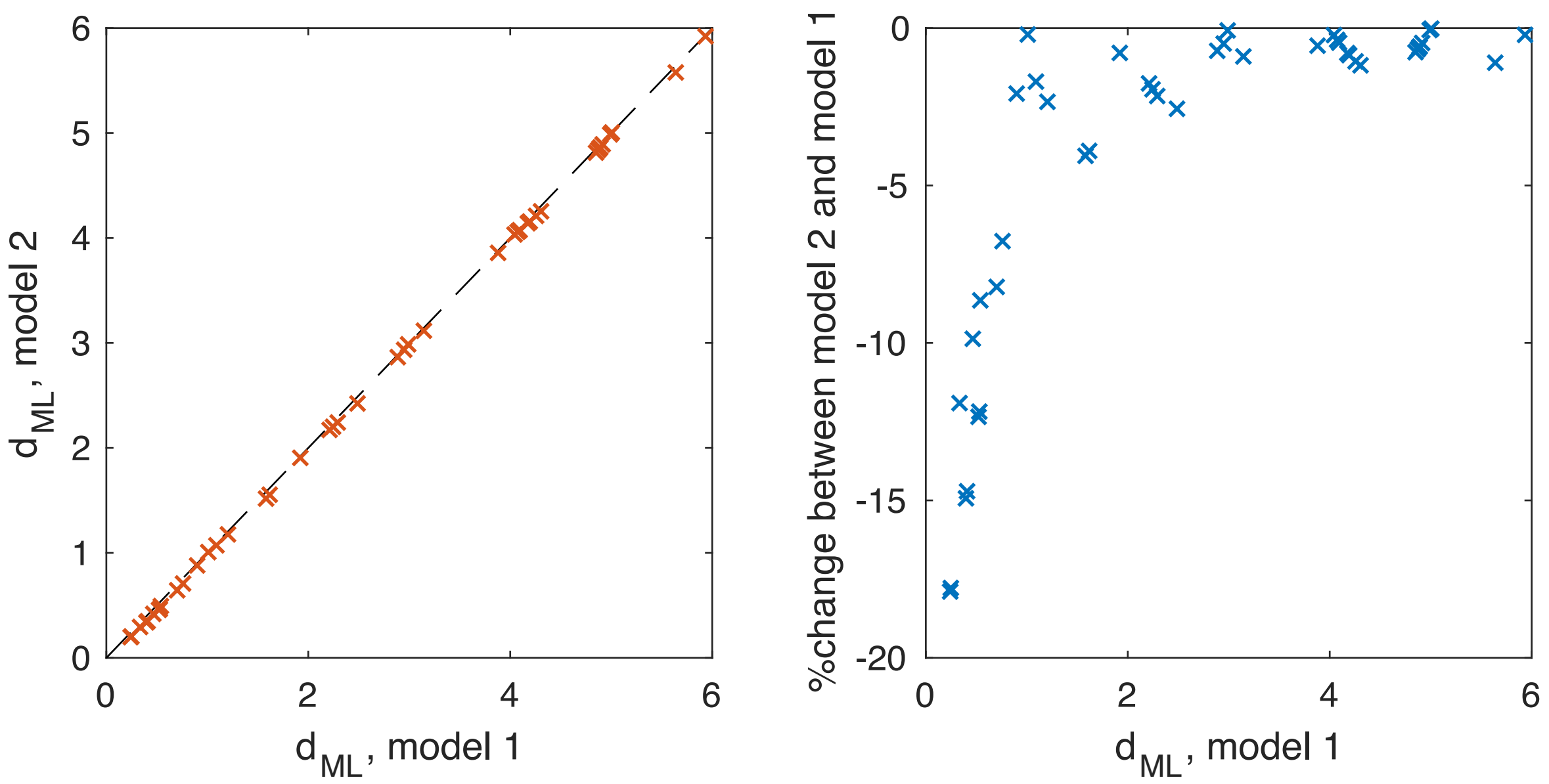

Figure S3: Comparison of the thickness in monolayers ($d_{ML}$) obtained from model 1 and model 2. Left: Values from model 2 plotted versus values from model 1 (red crosses), the black dashed line represents, where values would be equal. Right: Percentage change between model 2 and model 1 calculated as $(d_{ML}(model2)-d_{ML}(model1)/d_{ML}(model1)*100$.

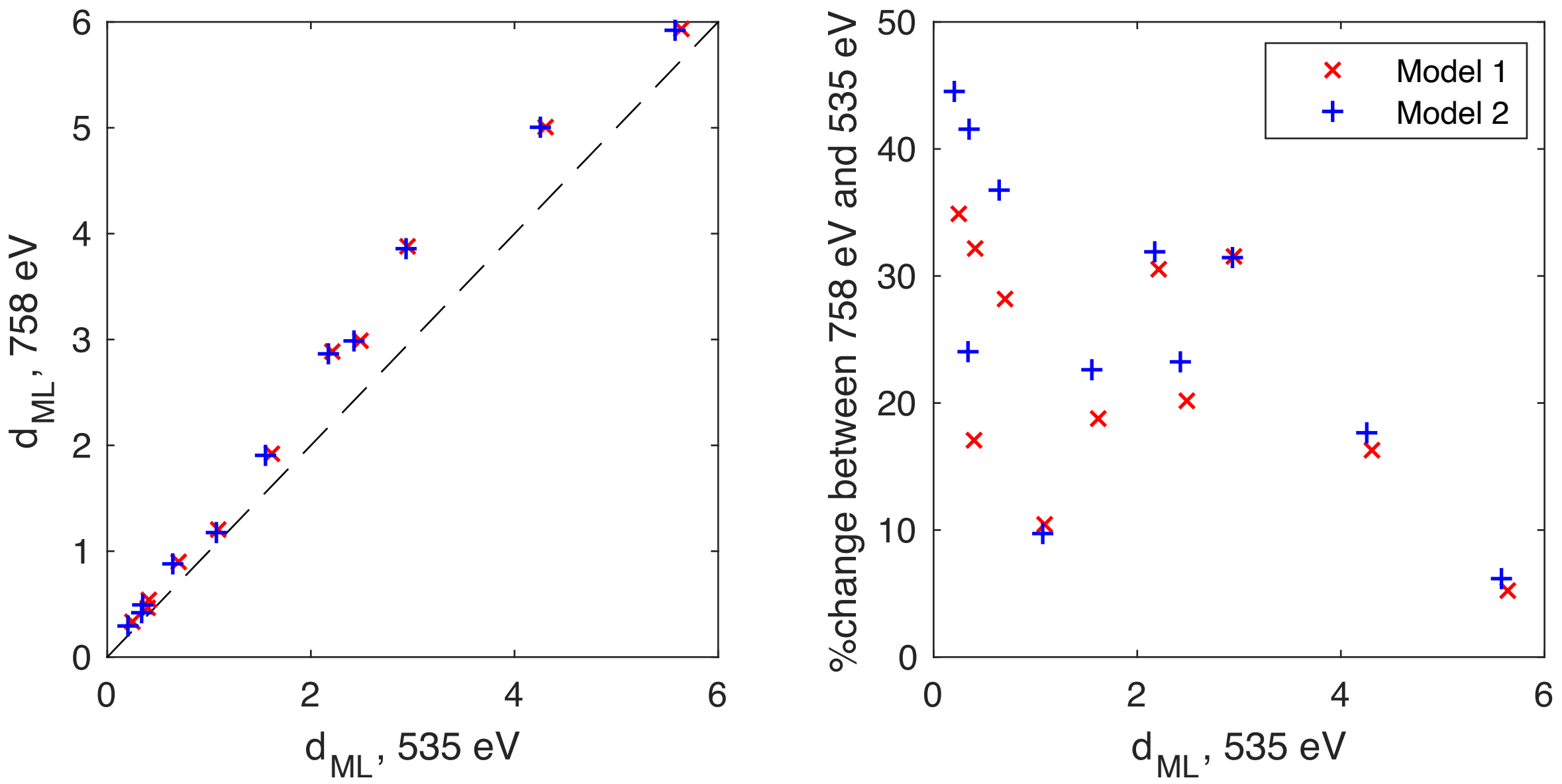

Figure S4: Comparison of the thickness in monolayers ($d_{ML}$) obtained from measurements with two different photon energies (535 eV and 758 eV) on the same spot. Left: Values from 758 eV plotted versus values from 535 eV (red crosses model 1, blue plusses model 2), the black dashed line represents where values would be equal. Right: Percentage change between values at 758 eV and 535 eV calculated as $((d_{ML}(758\ eV)-d_{ML}(535\ eV))/d_{ML}(535\ eV))*100$.

### Discussion of the ML thickness estimation

Model 1 and 2 are in good agreement in particular for higher coverage. At low coverage Model 2 is physically less meaningful and will therefore not give as good an estimate.

The differences in $d_{ML}$ estimated with the different photon energies can have several origins:

- experimental settings (X-ray intensity, sample position) not always precisely as assumed in the calculations);
- systematic error in the estimation of the mean free path (for example overestimating the relative difference between the mean free path at 535 eV and 758 eV);

- Systematic error in the model used, as it does not account for non-uniform layer formation.

However, it should be noted that the discrepancy between the values (20-30%) is small compared to the thickness range covered in the different experiments (< 1 ML coverage to > 5 ML). Overall, the discrepancy suggests that the error in the thickness estimation is about 30%. Model 1 is used to estimate thicknesses throughout the study.

### Additional photoelectron spectra

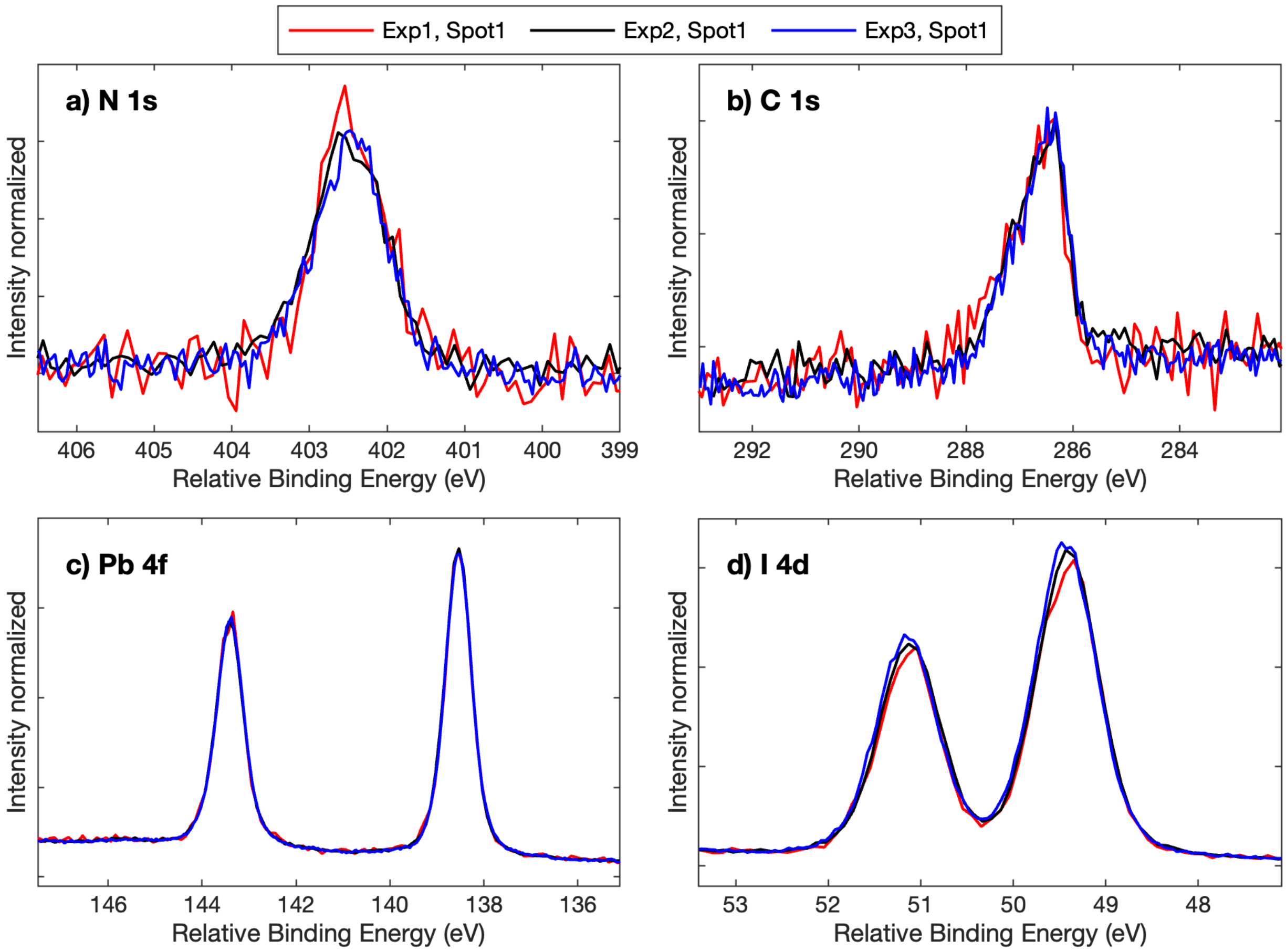

Figure S5: Normalised core level spectra of a) N1s, b) C 1s, c) Pb 4f, d) I 4d of $MAPbI_3$ single crystal surfaces after cleaving recorded with a photon energy of 758 eV. Data for three different single crystals measured on three different occasions (Exp1, Exp2 and Exp3) is shown. The binding energy scale is calibrated by setting the Pb $4f_{7/2}$ position to 138.54 eV and the intensities are normalised to the Pb 4f intensity.[5]

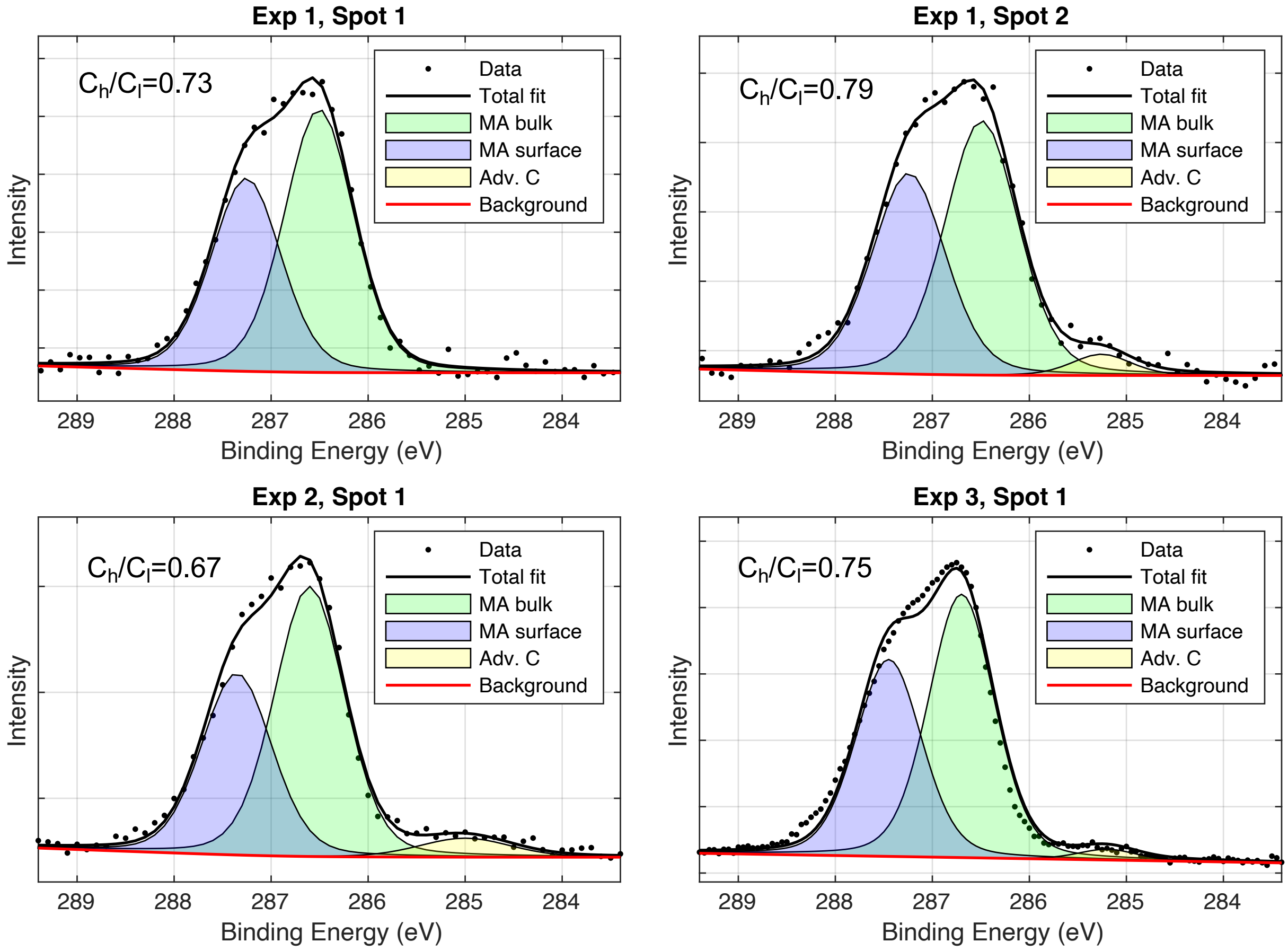

Figure S6: Curve fitting results of C 1s core level of $MAPbI_3$ single crystal surfaces after cleaving recorded with a photon energy of 535 eV. Data for three different single crystals measured on three different occasions (Exp1, Exp2 and Exp3) and two different measurements spots on one of the single crystals (Spot1, Spot2) is shown. The intensity ratios of the MA surface contribution (blue, $C_h$) and MA bulk contribution (green, $C_l$) are given in the figures. The binding energy scale is calibrated versus the Fermi level through measurement of the Au 4f core level on a gold foil.

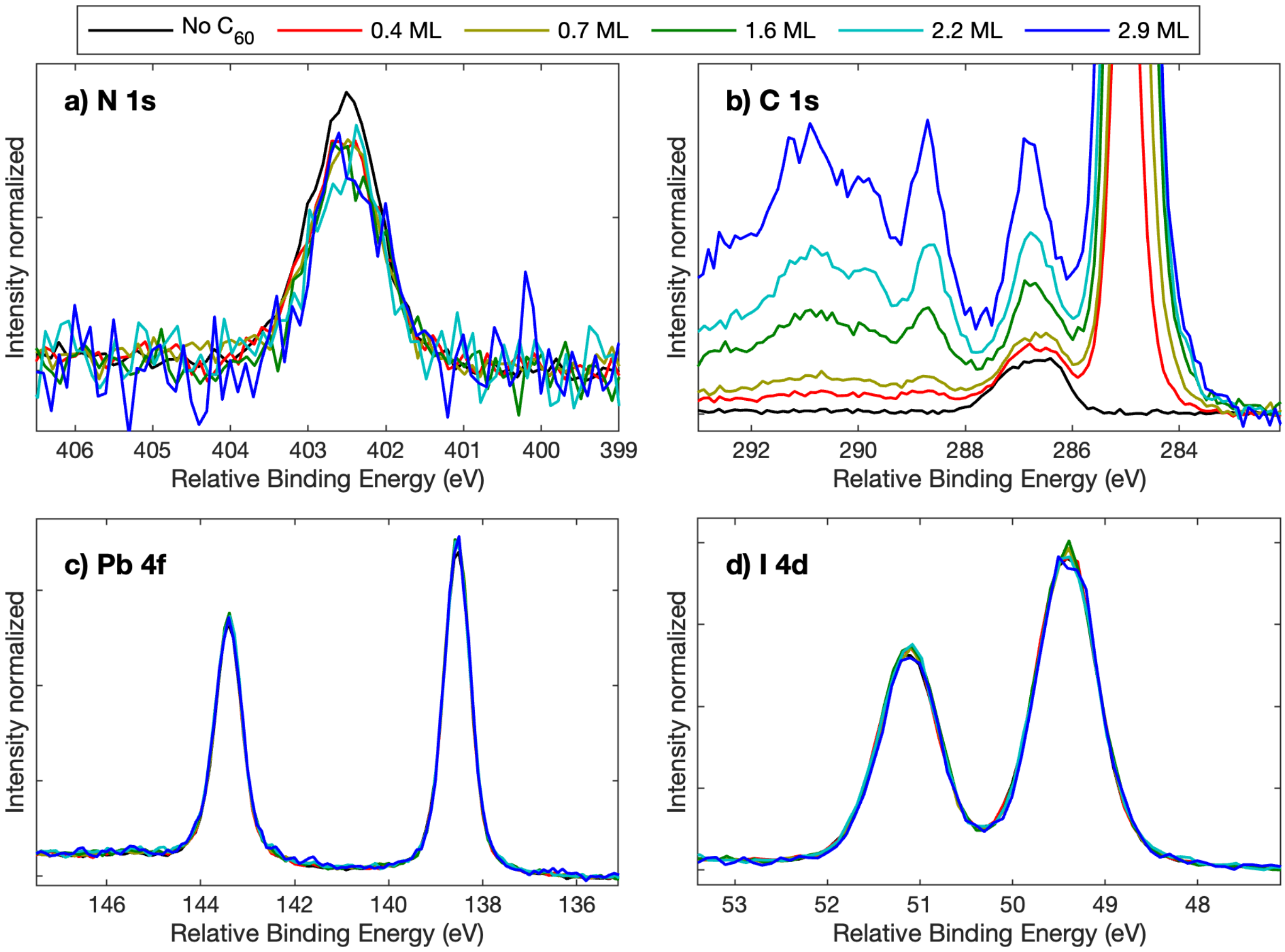

Figure S7. Normalized core level spectra of a) N1s, b) C 1s, c) Pb 4f, d) I 4d of a $MAPbI_3$ single crystal surface in Exp 1, spot 1 after cleaving and after consecutive evaporations of $C_{60}$ recorded with a photon energy of 535 eV. The binding energy scale is calibrated by setting the Pb $4f_{7/2}$ position to 138.54 eV and the intensities are normalized to the Pb 4f intensity. The y-axis of the C 1s spectra is zoomed in to focus on the perovskite peaks and C60 satellite peaks.

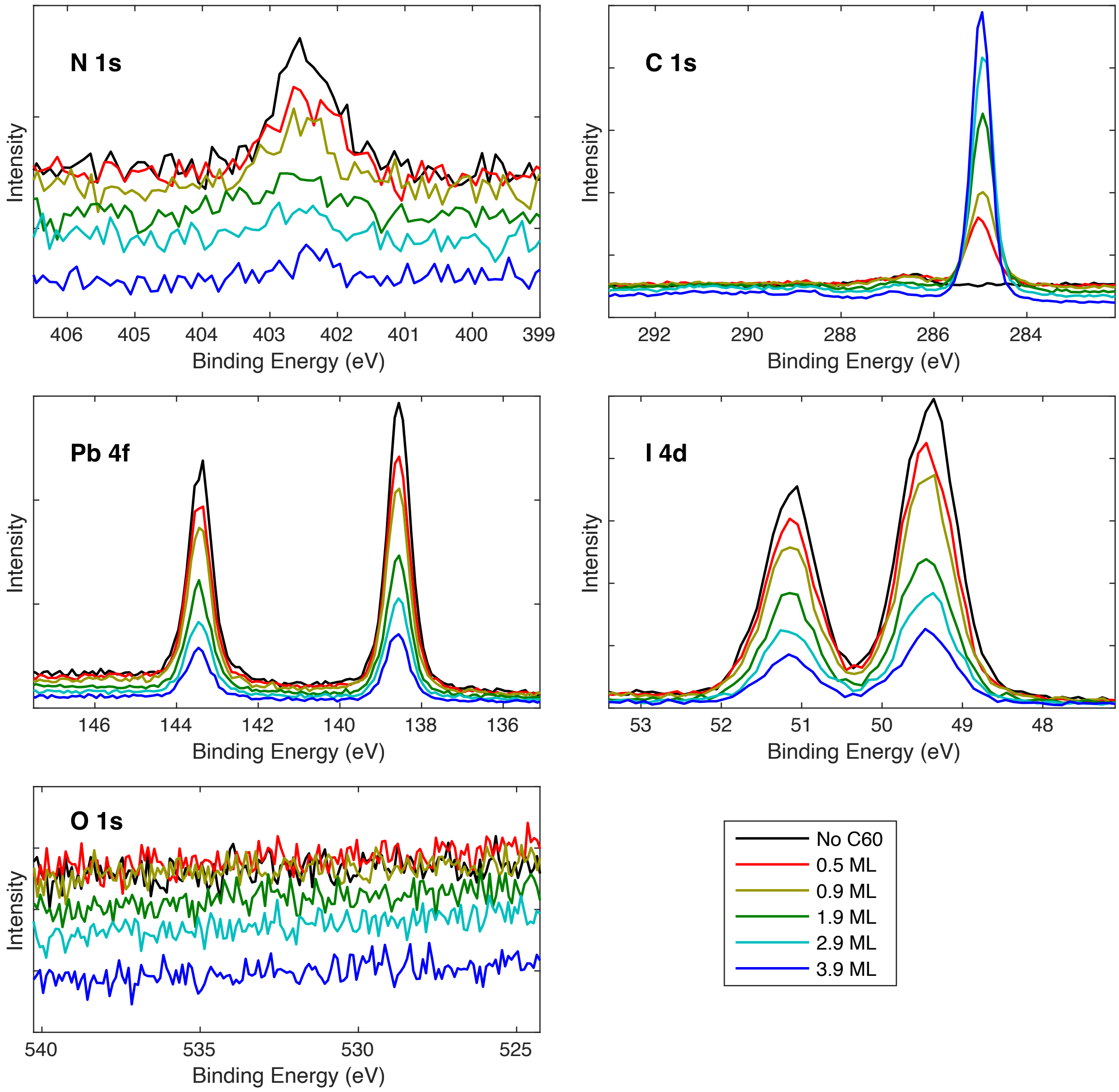

Figure S8. Core level spectra measured in Experiment 1, spot 1 after evaporation with an incident photon energy of 758 eV.

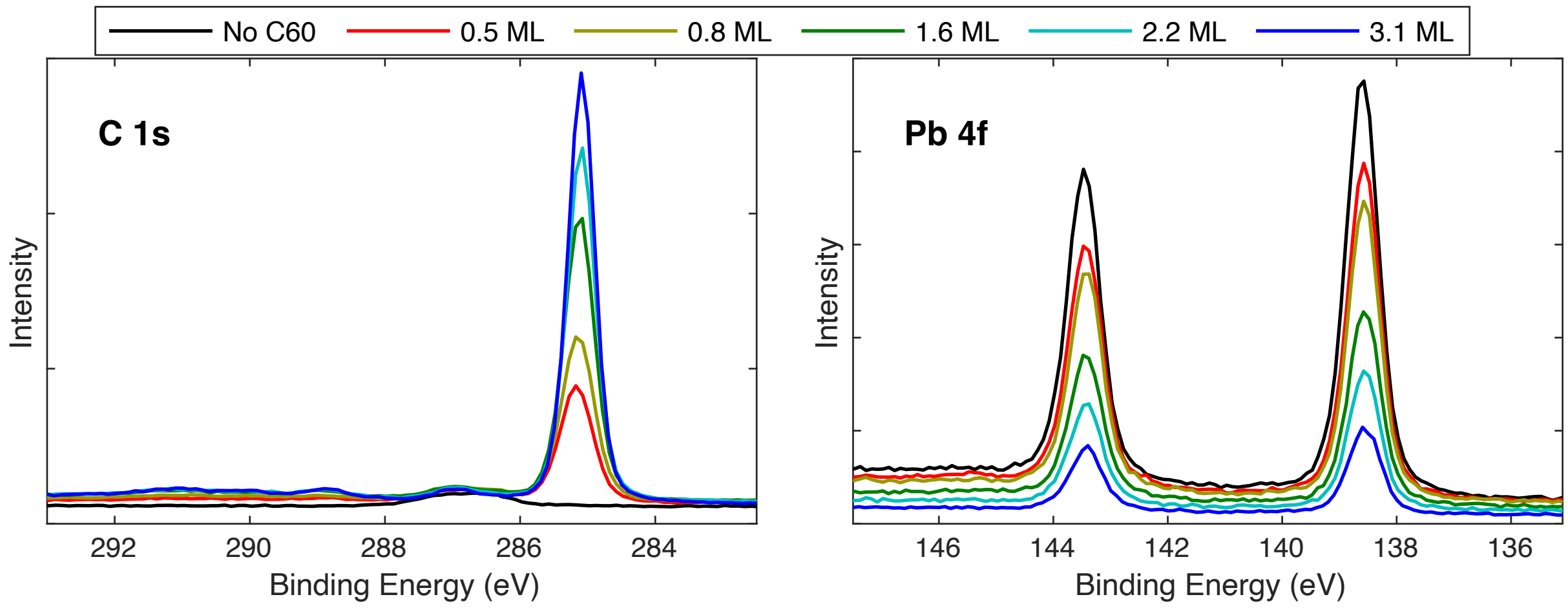

Figure S9. Core level spectra measured in Experiment 1, spot 2 after evaporation with an incident photon energy of 535 eV.

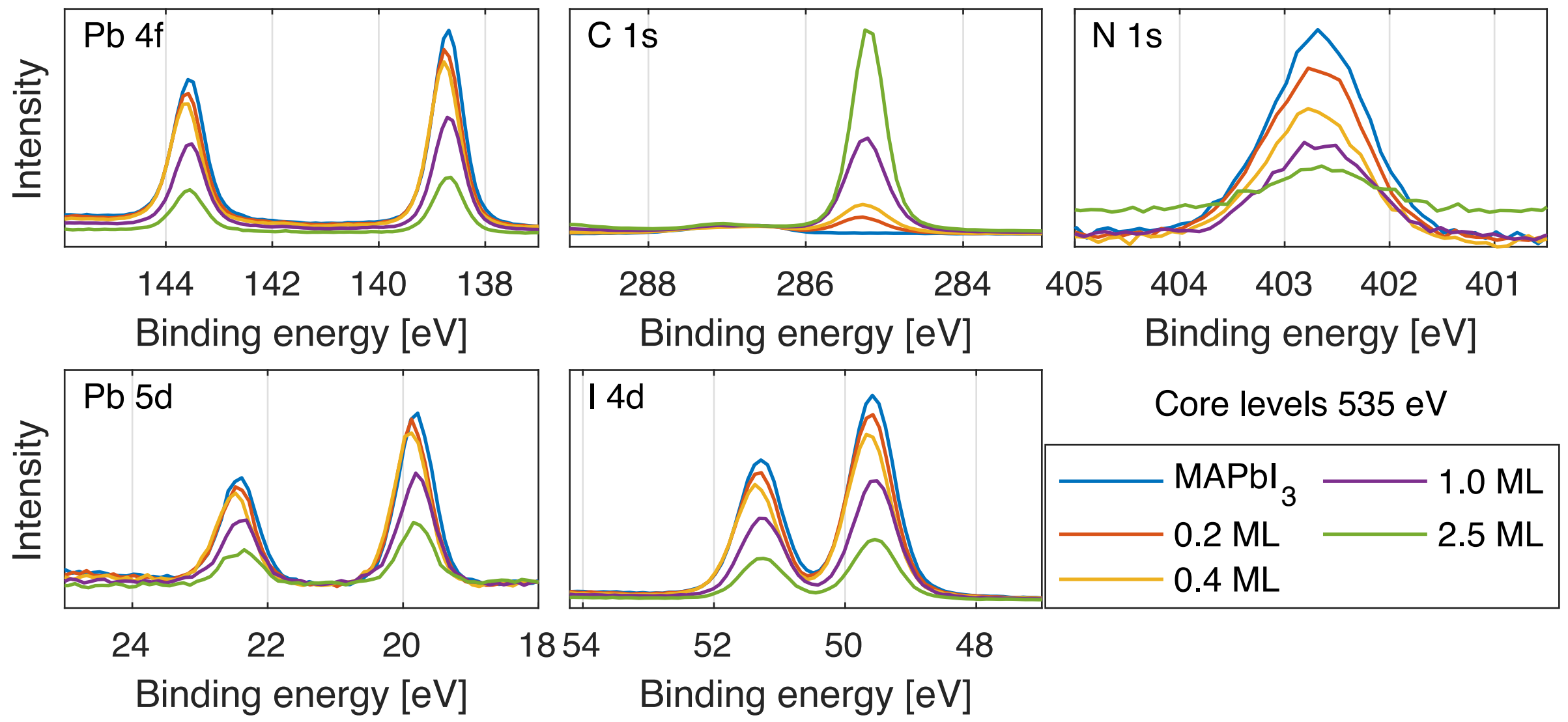

Figure S10. Core level spectra measured in Experiment 2, spot 1 after evaporation with an incident photon energy of 535 eV.

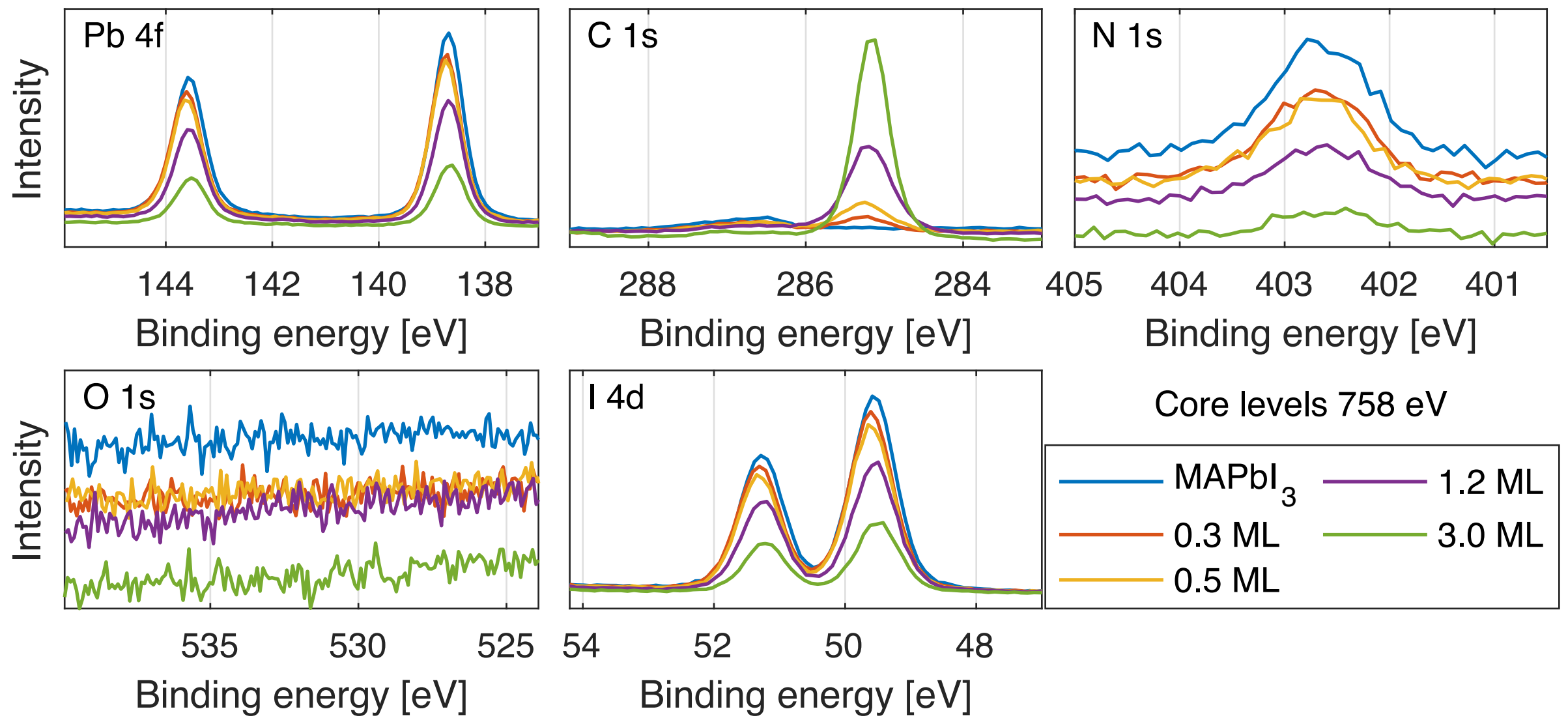

Figure S11. Core level spectra measured in Experiment 2, spot 1 after evaporation with an incident photon energy of 758 eV.

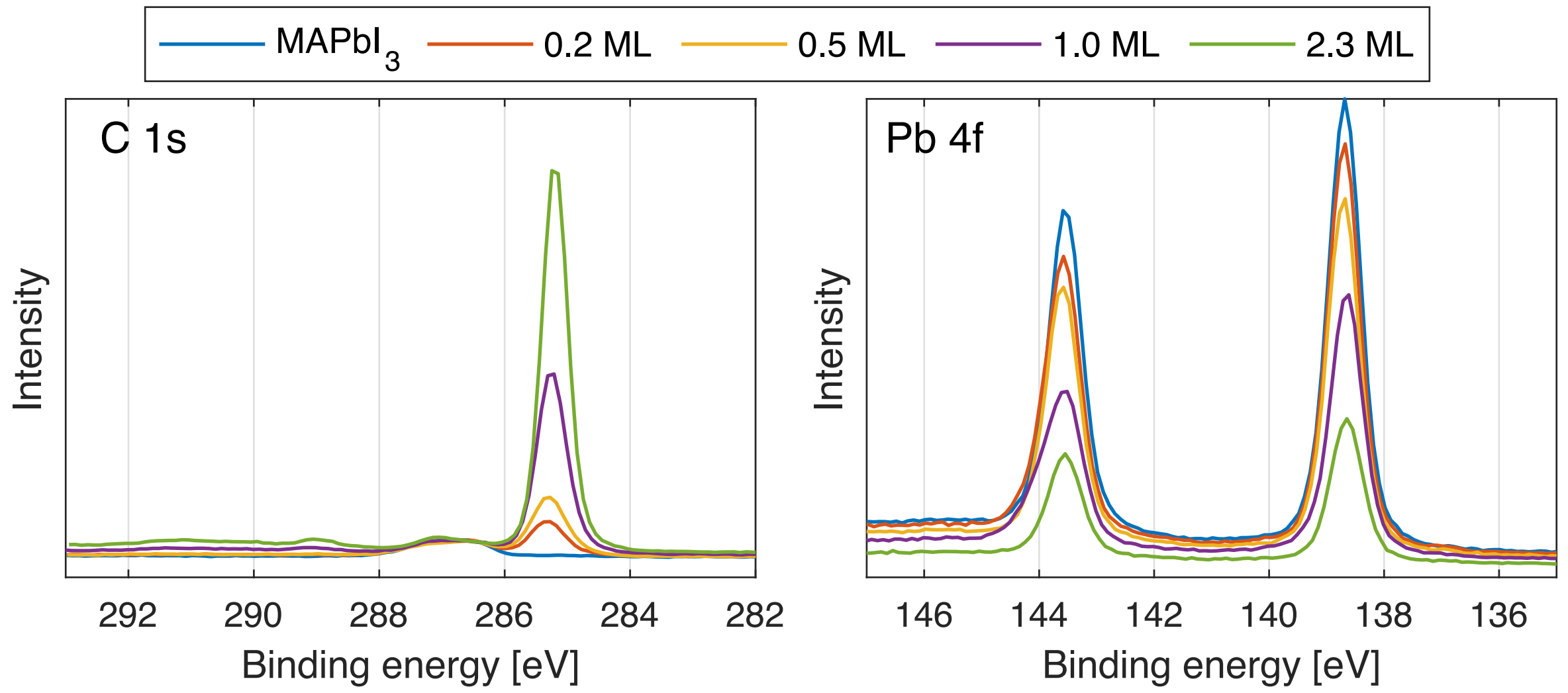

Figure S12. Core level spectra measured in Experiment 2, spot 2 after evaporation with an incident photon energy of 535 eV.

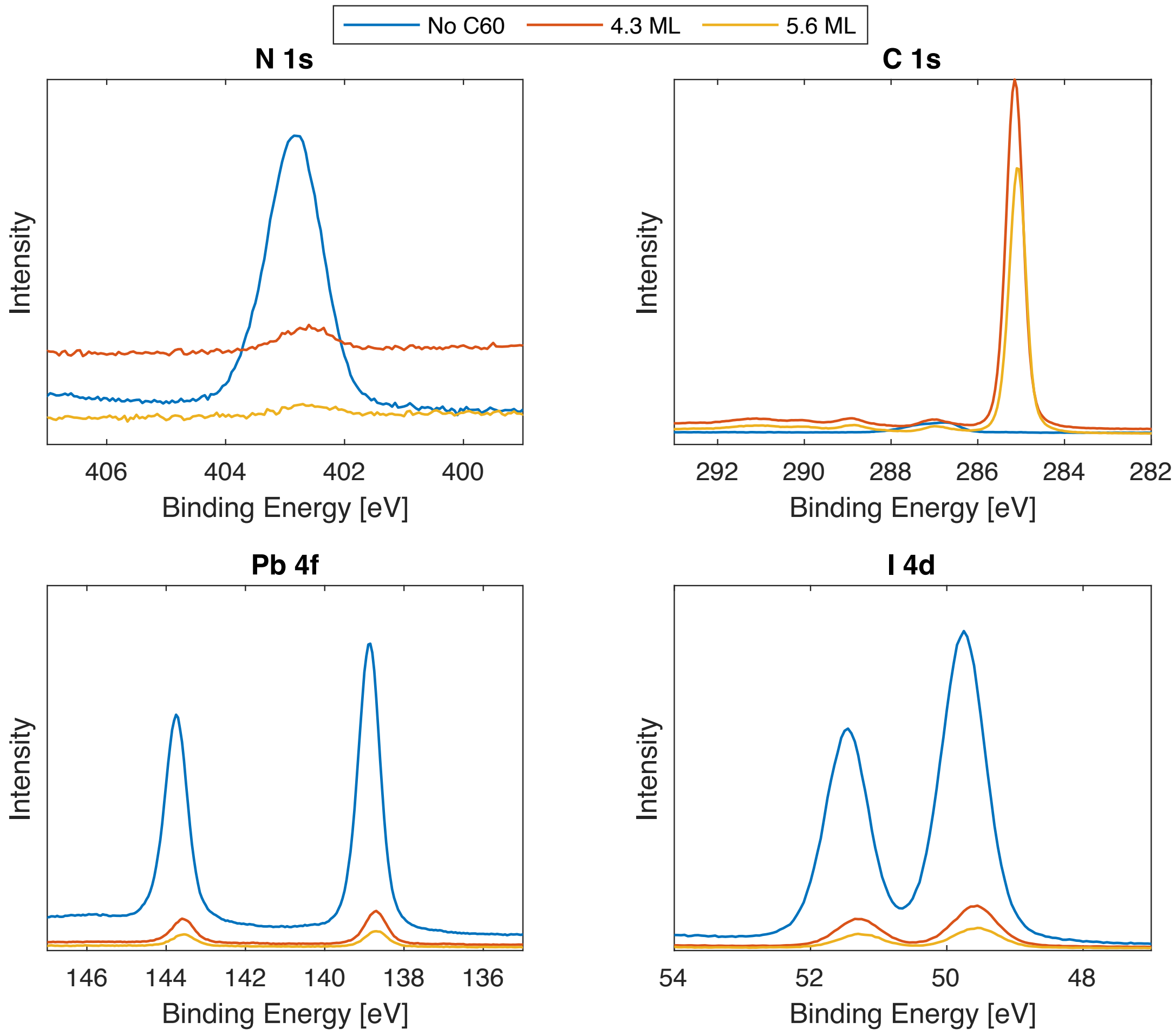

Figure S13. Core level spectra measured in Experiment 3, spot 1 after evaporation with an incident photon energy of 535 eV.

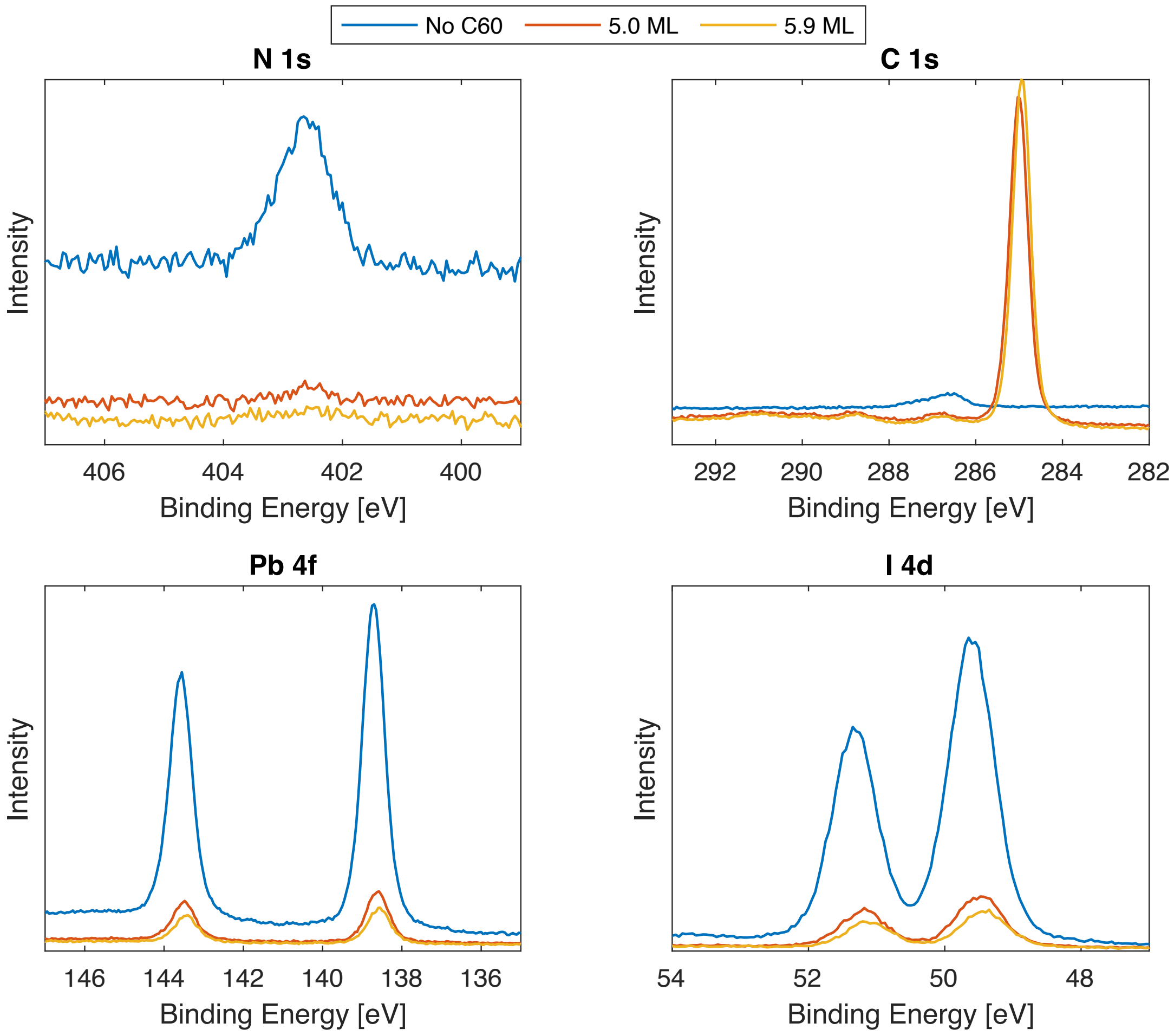

Figure S14. Core level spectra measured in Experiment 3, spot 1 with an incident photon energy of 758 eV.

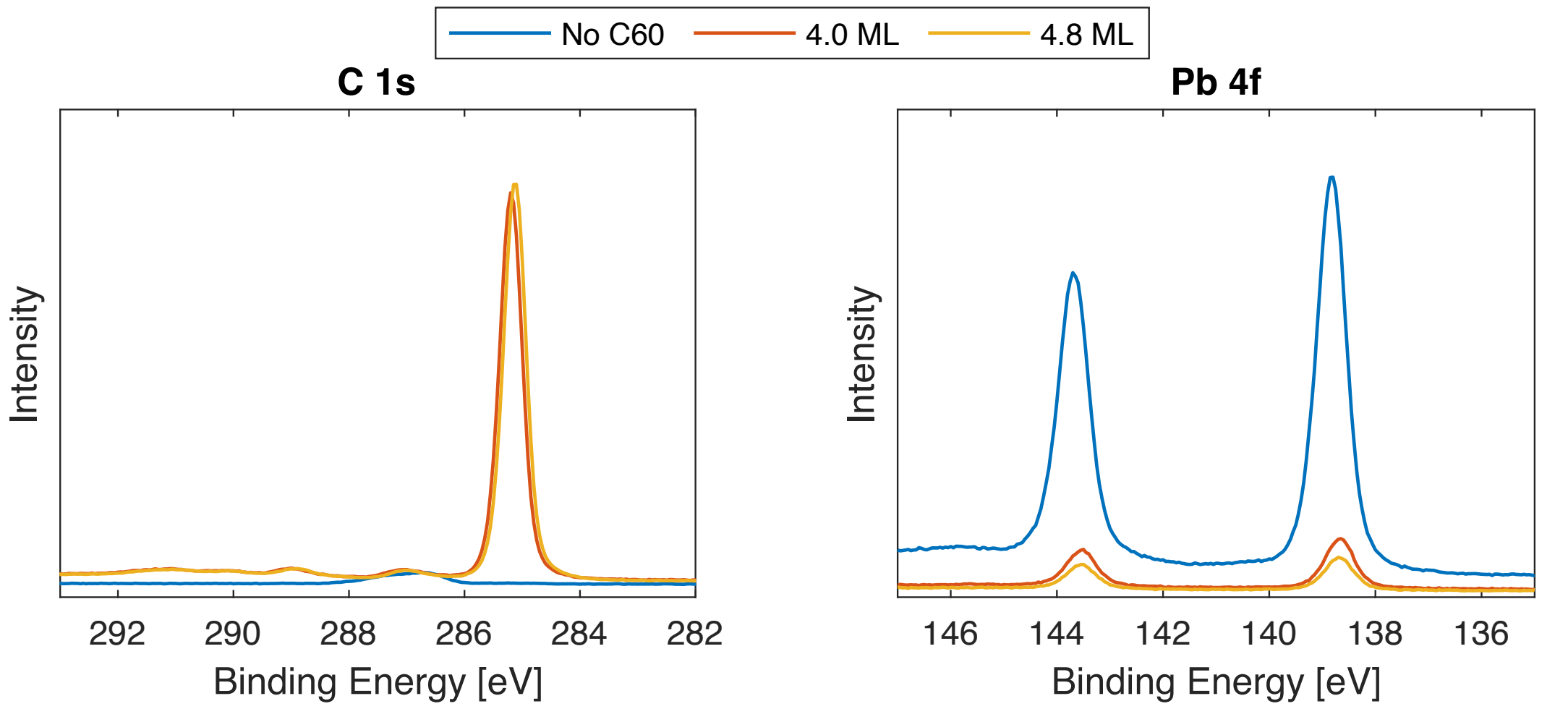

Figure S15. Core level spectra measured in Experiment 3, spot 2 with an incident photon energy of 535 eV.

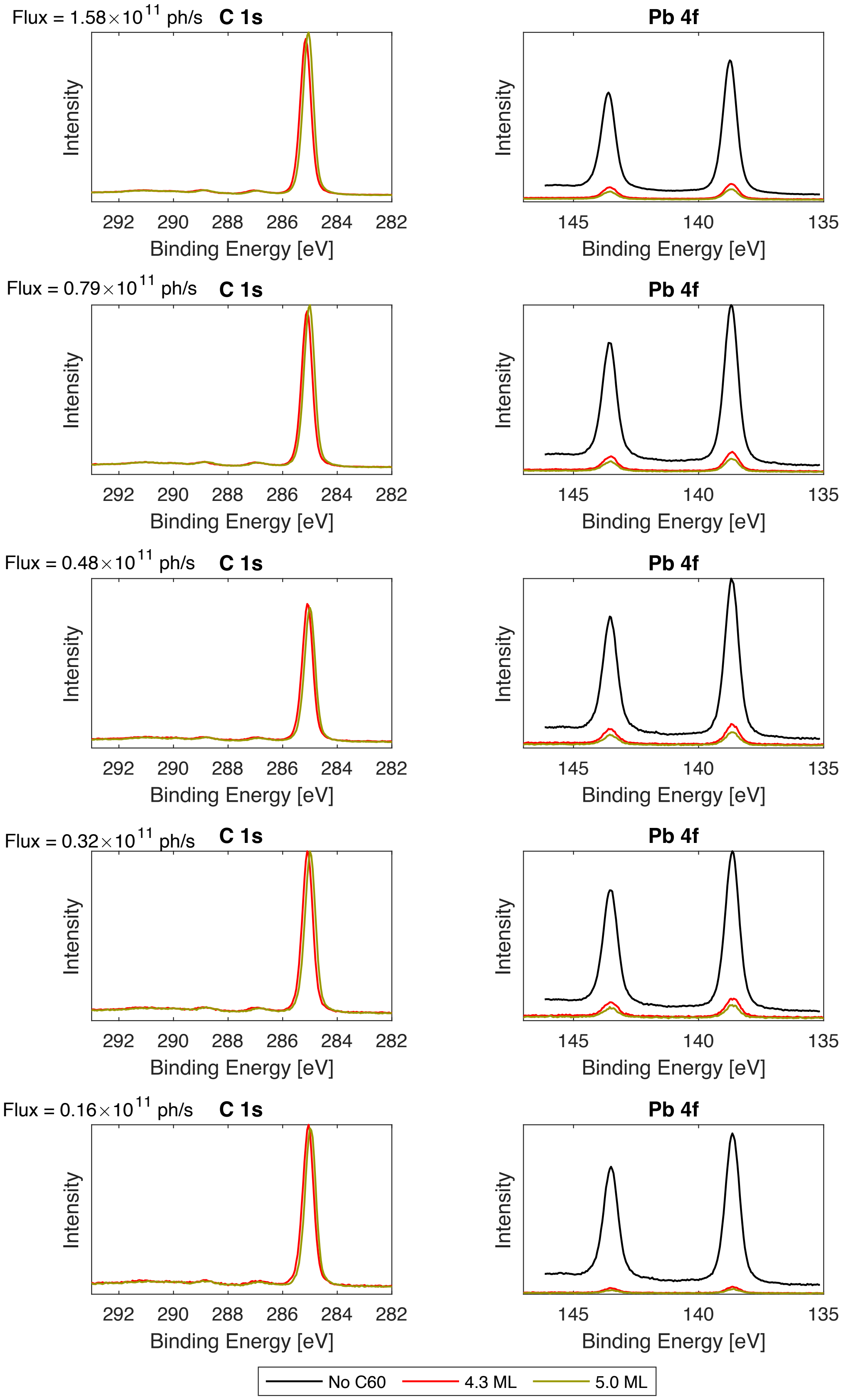

Figure S16. Core level spectra measured in Experiment 3, spot 3 at 535 eV incident photon energy with the photon flux density varied through adjusting the size of the exit slit.

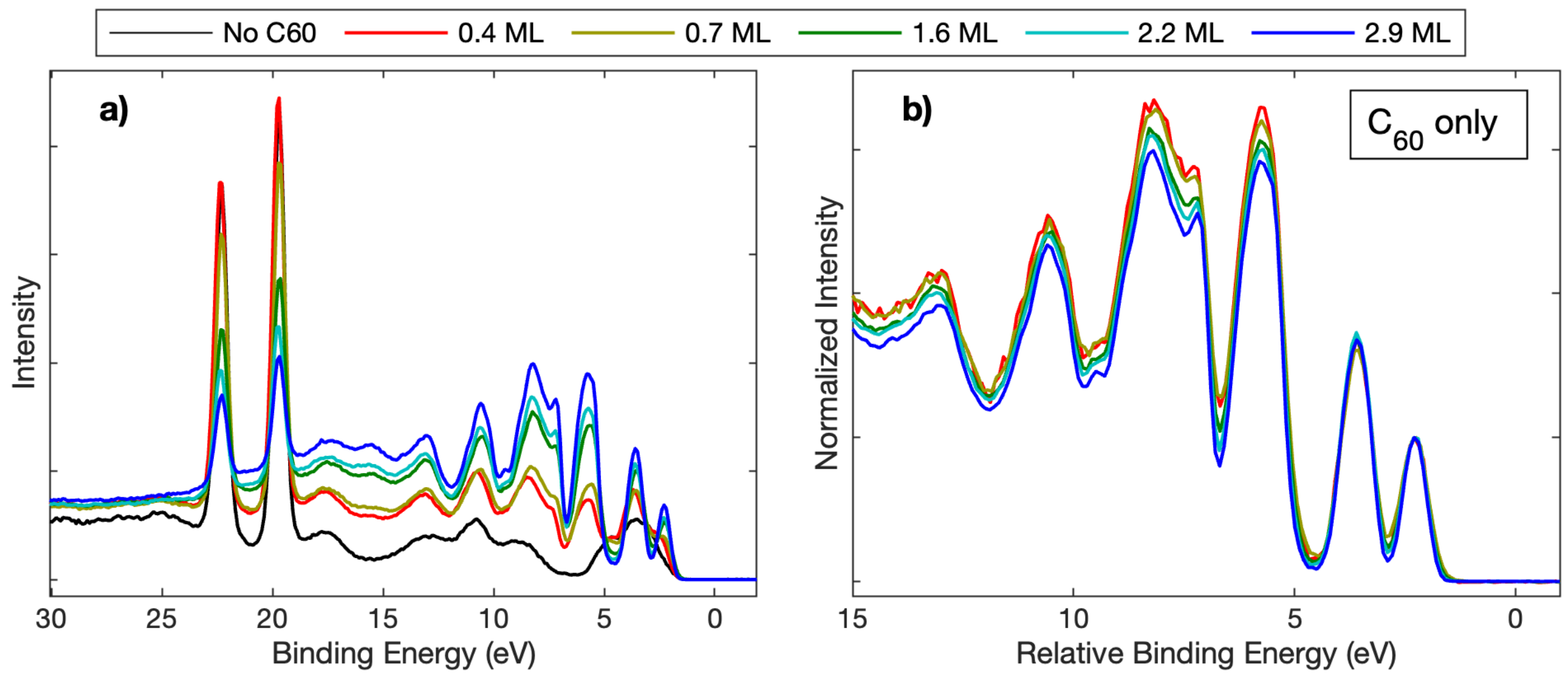

Figure S17: Valence band spectra measured in Exp 1, spot 1 after evaporation with an incident photon energy of 130 eV. a) The spectra as measured energy calibrated to the Fermi level of a gold reference foil. b) $C_{60}$-only spectra obtained by subtracting the contribution of the perovskite from the spectra after $C_{60}$ evaporation. These spectra are then energy aligned on the position of the HOMO peak of the $C_{60}$ and normalized to this peaks maximum intensity.

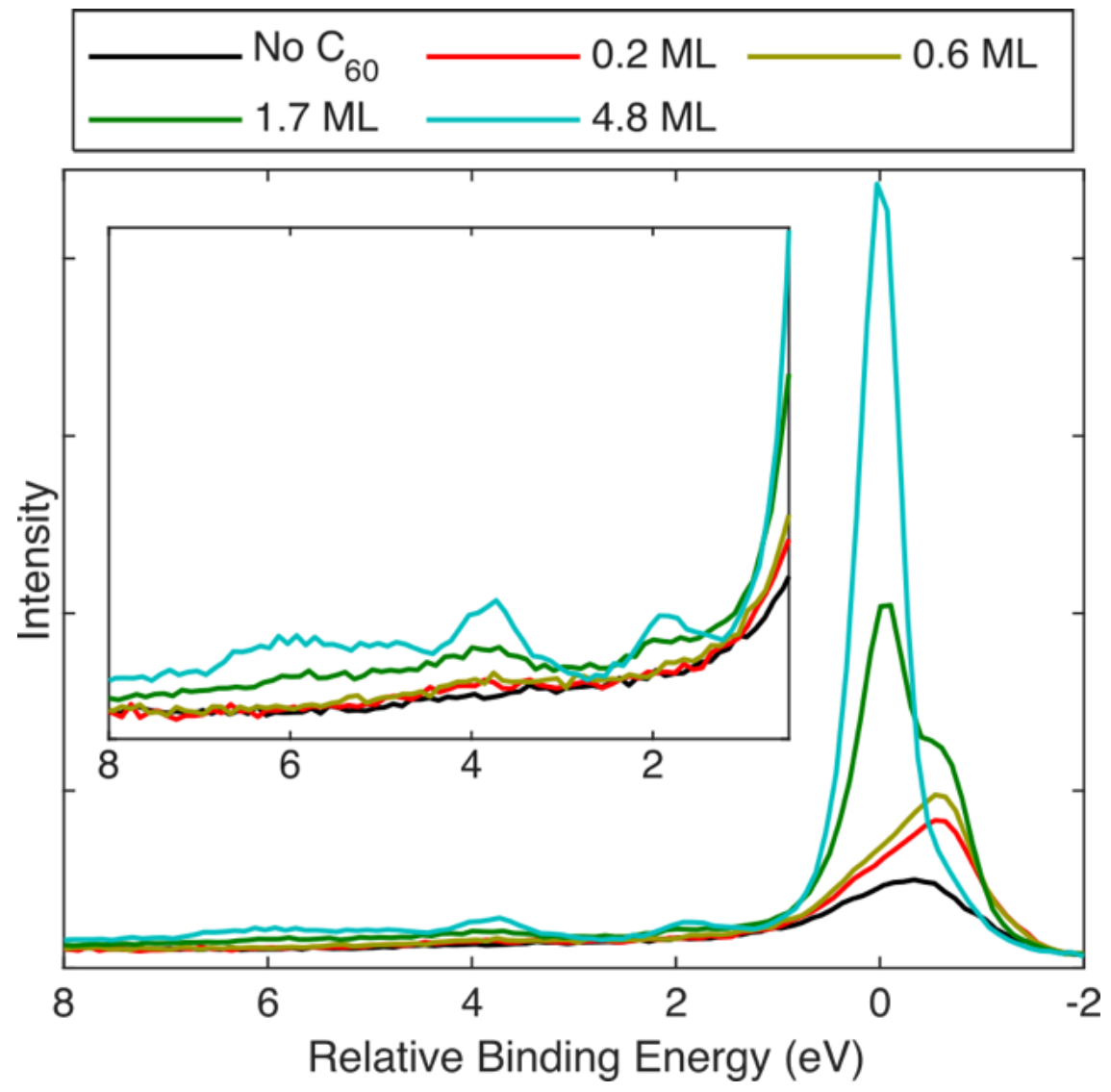

Fig S18 C 1s spectra of $C_{60}$ evaporated onto a gold foil in Exp 2 recorded with a photon energy of 535 eV relative to the position of the main $C_{60}$ peak in the 4.8 ML measurement. In the inset, a zoom on the region containing the $C_{60}$ satellite peaks is shown.

## Measurements at different photon flux values

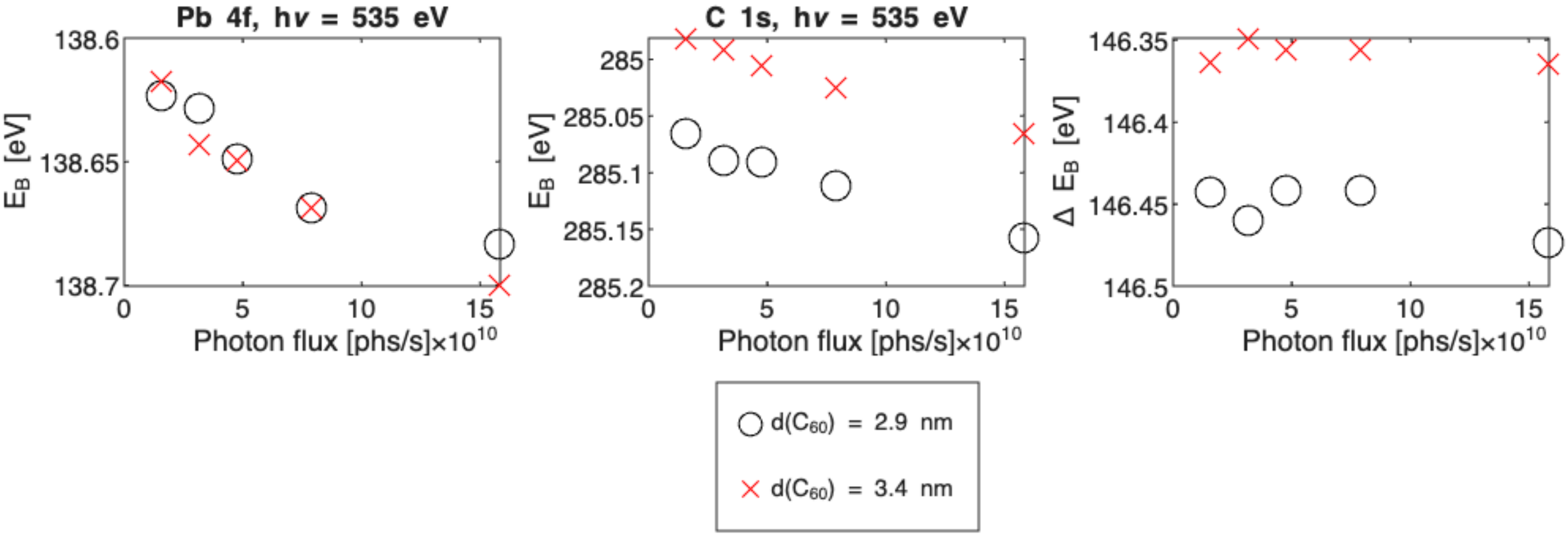

Figure S19 : Binding energy of the Pb $4f_{7/2}$ (left) and C 1s (middle) peak position and energy difference between the two (right) as a function of photon flux recorded in Exp3, spot 3, with 535 eV photon energy. The photon flux was varied by changing the size of the exit slit. The peak positions shift to higher binding energy with increasing photon flux, while the energy difference between them remains constant. This indicates reversible sample charging with higher X-ray intensities.

## Energy difference between Pb $4f_{7/2}$ peak and the $MAPbI_3$ valence band edge

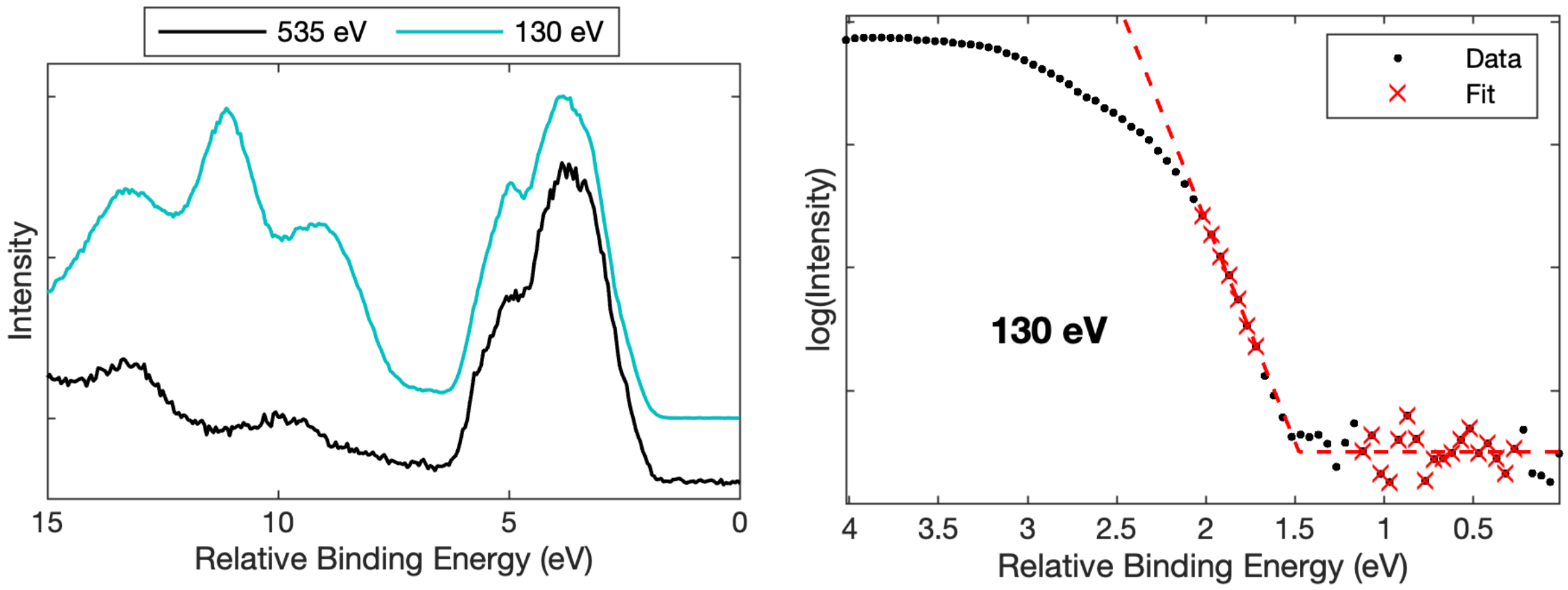

Figure S20: Left: Valence band spectra of MAPbI3 crystal measured in Exp 3, spot 1 with 2 photon energies (130 eV and 535 eV) aligned to the Pb $5d_{5/2}$ position at 20.01 eV recorded with a photon energy of 535 eV. Right: Valence band edge fit on a logarithmic axis to the data at 130 eV. The band edge was determined as the crossing point of the two fitted straight lines. The data at 130 eV was used due to the better signal-to-noise ratio at this photon energy.

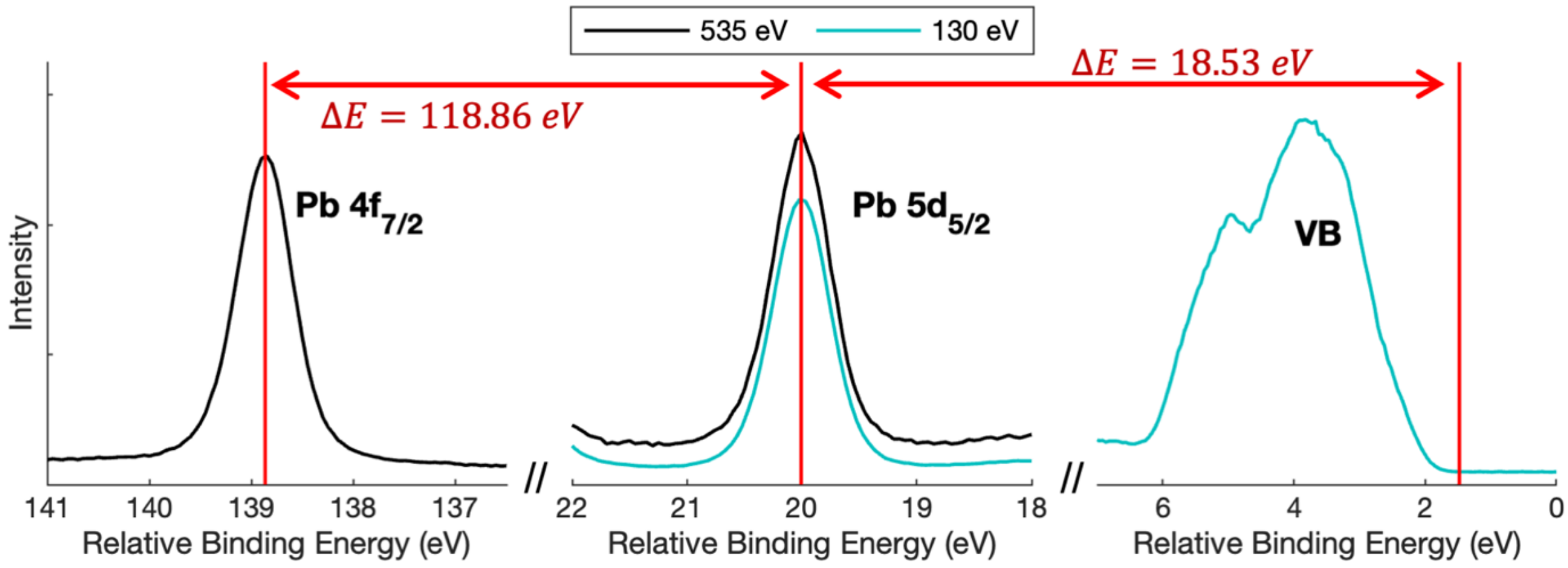

Figure S21: Spectra and positions recorded in Exp 3, spot 1 used for determining the energy offset between Pb $4f_{7/2}$ and the valence band edge in $MAPbI_3$. The energy difference between Pb $4f_{7/2}$ and Pb $5d_{5/2}$ (determined with a photon energy of 535 eV) was found to be 118.86 eV. The energy difference between Pb 5d5/2 and the valence band edge (determined with a photon energy of 130 eV) was found to be 18.53 eV. This gives an energy difference between Pb $4f_{7/2}$ and the valence band edge of 137.39 eV.

### Energy difference between C 1s peak and the $C_{60}$ HOMO edge

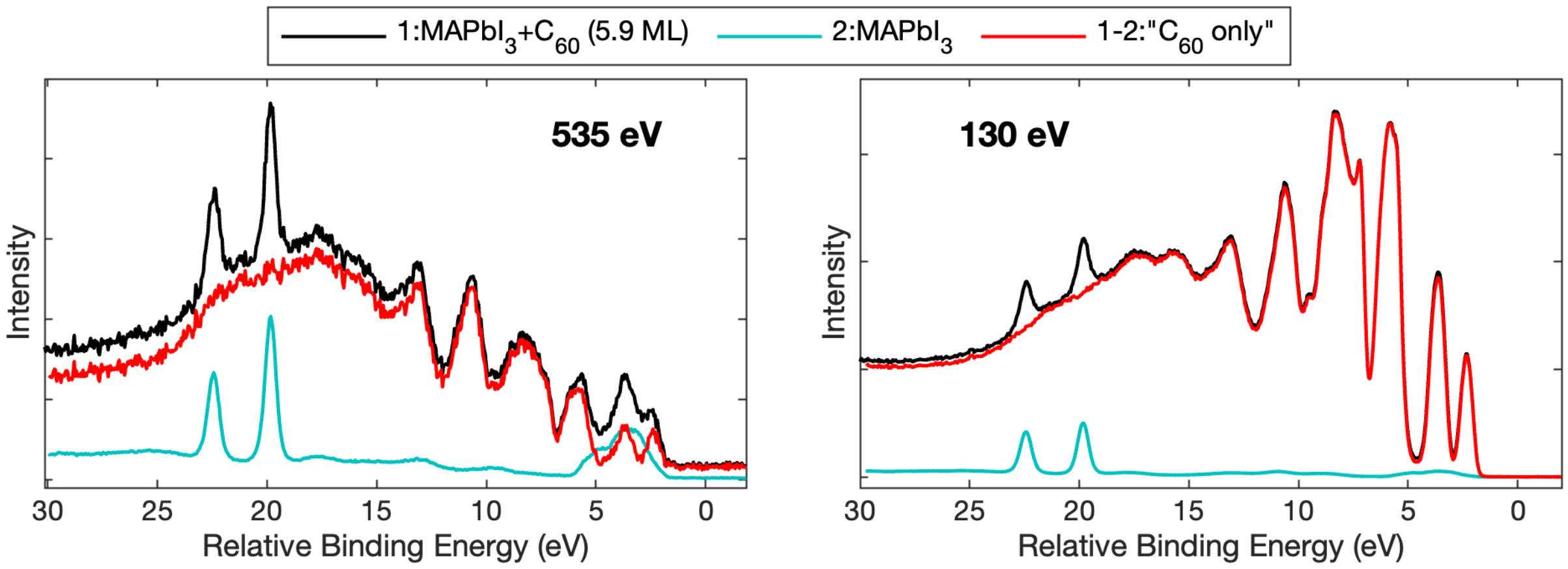

Figure S22: Valence band spectra of $C_{60}$ on $MAPbI_3$ crystal (5.9 ML) measured in Exp 3, spot 1 with two photon energies (535 eV, left and 130 eV,right) compared to spectra recorded on the clean $MAPbI_3$ surface in the same experiment. The $MAPbI_3$ spectra are calibrated and normalised to the Pb 5d intensity recorded in the spectra after evaporation. Using the two recorded spectra, $C_{60}$ only spectra are obtained by subtracting the $MAPbI_3$ contribution from the spectra after $C_{60}$ evaporation.

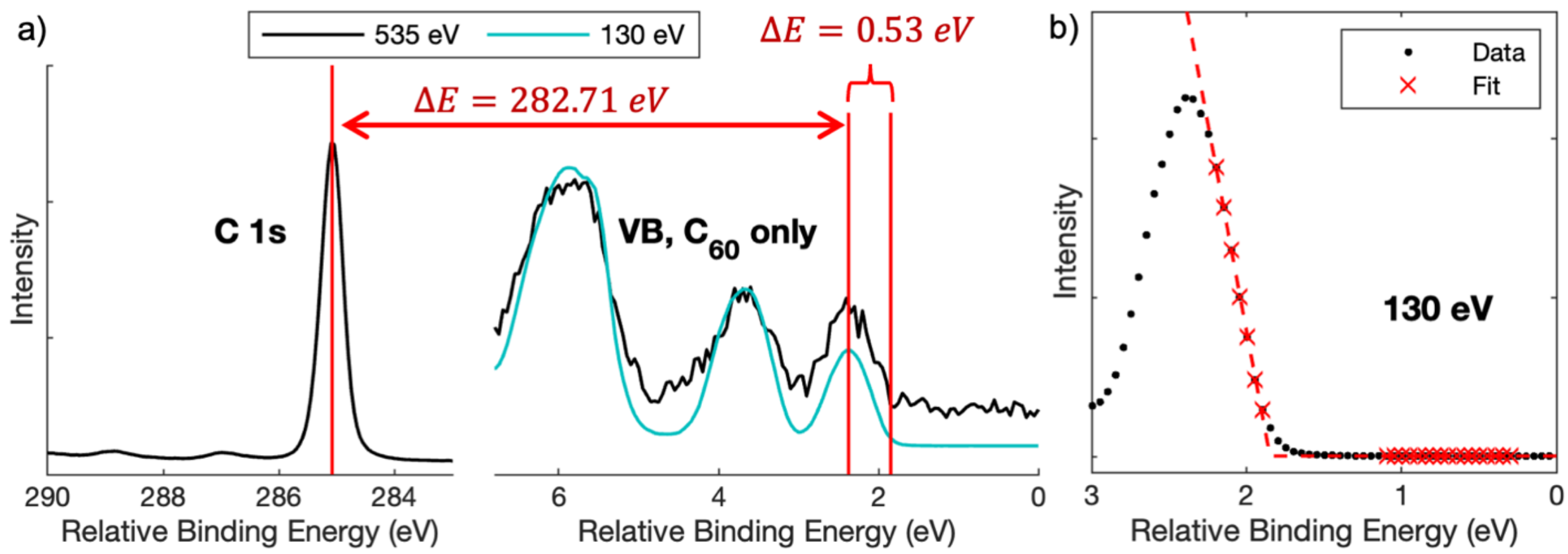

Figure S23: a) Spectra and positions recorded in Exp 3, spot 1 used for determining the energy offset between C 1s and the HOMO onset of $C_{60}$. The valence band spectrum at 130 eV is energy calibrated to the spectrum at 535 eV by placing the HOMO peak position at the same energy. The energy difference between C 1s and the HOMO peak position (determined with a photon energy of 535 eV) was found to be 282.71 eV. The energy difference between the HOMO peak and the HOMO onset energy (determined with a photon energy of 130 eV) was found to be 0.53 eV. This gives an energy difference between C 1s and the HOMO onset of 283.24 eV. b) Linear fit to the HOMO edge of the $C_{60}$ only spectrum recorded at 130 eV and calibrated to the spectrum at 535 eV used to determine the HOMO onset energy indicated in (a).